\documentclass[draft]{agujournal2019}
\usepackage{url} 
\usepackage{lineno}
\usepackage[inline]{trackchanges} 
\usepackage{soul}

\usepackage[version=3]{mhchem}
\usepackage{amssymb}
\usepackage{multirow} 
\usepackage{makecell}
\usepackage{ctable} 

\usepackage[italic]{hepnames}
\usepackage{threeparttable}

\draftfalse

\journalname{Reviews of Geophysics}

\begin{document}

\title{Neutrino Geoscience: Review, survey, future prospects}

\authors{ W. F. McDonough\affil{1,2,3}, H. Watanabe\affil{1}}

\affiliation{1}{Research Center for Neutrino Science, Tohoku University, Sendai, Miyagi, 980-8578, Japan}
\affiliation{2}{Department of Earth Sciences, Tohoku University, Sendai, Miyagi, 980-8578, Japan}
\affiliation{3}{Department of Geology, University of Maryland, College Park, MD 20742, USA} 

\correspondingauthor{William McDonough}{mcdonoug@umd.edu}

\begin{keypoints}
\item Parameters influencing the geological interpretation of a geoneutrino signal
\item Problems with establishing the mantle contribution to the geoneutrino signal
\item Importance of taking a geoneutrino flux measurement in the ocean far from the continents
\end{keypoints}
\bigskip

Author ORCID numbers are
 
     \begin{tabular}{rl}
William F. McDonough :  &  0000-0001-9154-3673   \\
Hiroko Watanabe :        &  0000-0002-2363-5637   \\
    \end{tabular}
 \bigskip

AGU Book chapter:\\
\hspace{2cm}\underline{Core-Mantle Coevolution: A Multidisciplinary approach}\\
\hspace{1.7cm} Editors: Takashi Nakagawa and Madhusoodhan Satish Kumar


\begin{abstract}
The earth's surface heat flux is 46$\pm$3 TW (terrawatts, 10$^{12}$ watts). Although many assume we know the earth's abundance and distribution of radioactive heat producing elements (i.e., U, Th, and K), estimates for the mantle's heat production varying by an order of magnitude and recent particle physics findings challenge our dominant paradigm. Geologists predict the earth's budget of radiogenic power at 20$\pm$10 TW, whereas particle physics experiments predict 15.3$^{+4.9}_{-4.9}$ TW (KamLAND, Japan) and 38.2$^{+13.6}_{-12.7}$ TW (Borexino, Italy). 

We welcome this opportunity to highlight the fundamentally important resource offered by the physics community and call attention to the shortcomings associated with the characterization of the geology of the earth. We review the findings from continent-based, physics experiments, the predictions from geology, and assess the degree of misfit between the physics measurements and predicted models of the continental lithosphere and underlying mantle. Because our knowledge of the continents is somewhat uncertain (7.1$^{+2.1}_{-1.6}$ TW), models for the radiogenic power in the mantle (3.5 to 32 TW) and the bulk silicate earth (crust plus mantle) continue to be uncertain by a factor of $\sim$10 and $\sim$4, respectively. Detection of a geoneutrino signal in the ocean, far from the influence of continents, offers the potential to resolve this tension. Neutrino geoscience is a powerful new tool to interrogate the composition of the continental crust and mantle and its structures.

\end{abstract}

\section*{Plain language summary}

Potassium, thorium, and uranium are radioactive and heat is given off during their decay. This release of energy drives the earth's dynamic processes of plate tectonics, mantle convection, and the geodynamo, with the latter generating the protective magnetic shield surrounding the planet. The earth is a hybrid vehicle powered by primordial and radiogenic energy, with the former from impacts during planet building and the latter from radioactive decays. Unfortunately, we lack a fuel gauge for either energy source and consequently we do not know how much power is left to drive the earth's engine, nor how much energy (and time) is left to keep it habitable?

Geoneutrinos are naturally occurring electron antineutrinos produced during $\beta^{-}$ (and $EC$, electron capture) decays of these heat producing elements. These tiny fundamental particles are almost impossible to detect, because they are about a billion times smaller than a proton, near-massless and chargeless. In 2005, particle physicists first detected the earth's emission of geoneutrinos with a large underground detector and are now telling us about the amount of radiogenic heat inside the earth \cite{araki2005experimental}. Reading the earth's fuel gauge by counting geoneutrinos, however, requires that geologists understand the abundance and distribution of these heat producing elements in the continents and mantle. 

New results from geoneutrino detectors in Japan and Italy present contrasting stories as to how much fuel is left in the earth's engine. Geologists are building 3-D physical and chemical models of the earth, but are vexed by the subsurface complexities of the continents. Efforts are underway to construct an ocean-going geoneutrino detector that can independently measure the mantle's flux, far from the influence of continents, and define how much radiogenic fuel is left in the earth.

\section{Introduction}
Core-mantle evolution involves understanding earth's differentiation processes, which established the present day distribution of the heat producing elements, and its dynamic consequences (i.e., the radiogenic heat left in the mantle powering mantle convection, plate tectonics, and the geodynamo).  The energy to drive the earth's engine comes from two different sources: primordial and radiogenic. Primordial energy represents the kinetic energy inherited during accretion and core formation. Radiogenic energy is the heat of reaction from nuclear decay. We do not have a constraint on the proportion of these different energy sources driving the present day earth's dynamics. In turn, this means that we do not have sufficient constraint on the thermal evolution of the planet, aside from first order generalities. You might ask, is this important? We ask the question -- how much energy (and time) is left to keep the earth habitable?

We understand that the earth started out hot due to abundant accretion energy, the gravitational energy of sinking metal into the center, a giant impact event for the formation of the earth's Moon, and energy from short-lived (e.g., $^{26}$Al and $^{60}$Fe) and long-lived (K, Th, and U) radionuclides. From this hot start the planet should quickly lose some of its initial energy, although the amount and rate are unknowns. There are many significant unknowns regarding the thermal evolution of the earth: (1) the nature and presence (or absence) and lifetime of an early atmosphere, which has a thermal blanketing effect; (2) the compositional model for the earth, particularly the absolute abundances of the heat-producing elements (HPE: K, Th, \& U); (3) the cooling rate of the mantle (present day estimates: 100$\pm$50 K/Ga); and (4) the rate of crust formation and thus extraction of HPEs from the mantle. 

The recent recognition \cite{KRAUSS} and subsequent detection \cite{araki2005experimental} of the planet's geoneutrino emission has opened up a new window into global scale geochemistry of the present day earth. The measurement of this flux presents earth scientists with a transformative opportunity for new insights into the composition of the earth and its energy budget.  For the most part, solid earth geophysics measures and parameterizes the present day state of the planet. In contrast, solid earth geochemistry measures and parameterizes its time integrated state, mostly on a hand sample scale and then extrapolates these insights to larger scales. The advent of measuring the earth's geoneutrino flux allows us, for the first time, to get a global measure of its present day amount of Th and U.

This paper is organized as follows: we provide the rationale for the field of neutrino geoscience and define some terms (Section 2). We review the existing and developing detectors, the present-day detection methods and future technologies (Section 3). We discuss the latest results from the physics experiments (Section 4). We present the range of compositional models proposed for the earth (Section 5) followed by a discussion of the geological prediction of the geoneutrino fluxes at various detectors (Section 6). We finish with a discussion on determining the radioactive power in the mantle (Section 7) and future prospects (Section 8).

\section{Neutrino Geoscience}
The field of neutrino geoscience focuses on constraining the earth's abundances of Th and U and with these data we can determine: (1) the absolute concentration of refractory elements in the earth and from that determine the bulk silicate earth's composition (BSE; crust plus mantle), and (2) the amount of radiogenic power in the earth driving the planet's major dynamic processes  (e.g., mantle convection, plate tectonics, magmatism, and the geodynamo). These two constraints set limits on the permissible models for the composition of the earth and its thermal evolutionary history.

First, the \textit{refractory elements} are in constant relative abundances in all chondrites. There are 36 of these elements (e.g., Al, Ca, Sr, Zr, REE, Th \& U) and by establishing the absolute abundance of one defines all abundances, since refractory elements exist in constant ratios to each other \cite{McDonough1995}. Most of these elements are concentrated in the bulk silicate earth, but not all (e.g., Mo, W, Ir, Os, Re, etc) and these latter ones are mostly concentrated in the metallic core. Knowing the earth's abundance of Ca and Al, two of the eight most abundant elements (i.e., O, Fe, Mg, Si, Ca, Al, Ni, and S) that make up terrestrial planets (i.e., 99\%, mass and atomic proportions), defines and restricts the range of accepted compositional models of the bulk earth and BSE.

Second, the the decay of $^{40}$K, $^{232}$Th, $^{238}$U, and $^{235}$U (i.e., HPE) provides the earth's radiogenic power, accounts for 99.5\% of its total radiogenic power, and is estimated to be ($19.9\pm3.0$) TW (1 TW = 10$^{12}$ watts). This estimate, however, assumes a specific BSE model composition \cite{McDonough1995,palme2014cosmochemical}. It must be noted that there is no consensus on the composition of the BSE, and so predictions from competing compositional models span from about 10 to 38 TW \cite{javoy2010chemical,agostini:2020}. This uncertainty in our present state of knowledge means that the field of neutrino geoscience plays a crucial role in resolving fundamental questions in earth sciences.

\subsection{Background Terms}
The field of neutrino geoscience spans the disciplines of particle physics and geoscience, including geochemistry and geophysics. The following list of terms are offered to support this interdisciplinary research field.

\noindent \textit{alpha ($\alpha$) decay}: a radioactive decay process that reduces the original nuclide ($X$) by 4 atomic mass units by the emission of a $^4_2$He nucleus and reaction energy ($Q$). Commonly, the $\alpha$ particle is emitted with between 4 and 9 MeV (1 MeV = 10$^6$ eV) of discrete kinetic energy. The basic form of $\alpha$ decay is:
\begin{equation} 
\begin{split} 
\textrm{Alpha} \quad (\alpha)\qquad   &^A_ZX \rightarrow \;^{A-4}_{Z-2}X' + \alpha +Q \\
\end{split}
\end{equation}

\noindent \textit{beta decay} : a radioactive decay process that transforms the original nuclide ($X$) into an isobar (same mass $A$) with the next lower proton number ($Z$) during either electron capture ($EC$) or $\beta^+$ decays or alternatively, the next higher proton number ($Z$) during $\beta^-$ decay. During each decay, there is an exchange of two energetic leptons (i.e., beta particles) and reaction energy ($Q$). Basic forms are: \newline
\begin{equation} 
\begin{split} 
\textrm{Beta Minus} \quad (\beta^-) \qquad &^A_ZX \rightarrow^A_{Z+1}X' + e^- + \APnue + Q, \\
\textrm{Electron Capture} \quad (\textrm{$EC$}) \qquad  &^A_ZX + e^- \rightarrow ^A_{Z-1}X' + \Pnue + Q, \\
\textrm{Beta Plus} \quad (\beta^+) \qquad &^A_ZX \rightarrow^A_{Z-1}X' + e^+ + \Pnue + Q \\
\end{split}
\end{equation}

\noindent \textit{beta particles} : First generation energetic leptons, either matter leptons (electrons and neutrinos: $e^-$ and $\Pnue$) or antimatter leptons (positrons and antineutrinos: $e^+$ and $\APnue$).

\noindent \textit{chondrite}: an undifferentiated stony meteorite containing chondrules (flash-melted spheres, sub-mm to several mm across), matrix (fine grained (micron scale) aggregate of dust and crystals), and sometimes Ca-Al-inclusions and other refractory phases. They are typically mixtures of silicates and varying amounts of Fe-Ni alloys and classified into groups based on their mineralogy, texture, and redox state. Three dominant groups are the carbonaceous, ordinary, and enstatite type chondrites, from most oxidized to reduced, respectively. Isotopic observations are also used to create a twofold classification of chondrites and related meteorites (i.e., the NC and CC groups). The NC (non-carbonaceous) group include enstatite and ordinary chondrites and are believed to have formed in the inner solar system inside of Jupiter. The CC (carbonaceous) group include carbonaceous chondrites and are believed to have formed in the outer solar system from Jupiter and beyond. The CI carbonaceous chondrite type (the sole chondrite type lacking chondrules) is considered most primitive because its element abundances matches that of the solar photosphere 1:1 over 6 orders of magnitude, except for the noble and H-C-N-O gases. 

\noindent \textit{geoneutrinos}: naturally occurring electron antineutrinos ($\APnue$, with the over-bar indicating it is an antimatter particle), mostly, and electron neutrinos ($\Pnue$), much less so, produced during $\beta^-$, and [$EC$ and $\beta^+$] decays, respectively. The interaction cross-sections, which scale with their energy, for the detectable geoneutrinos (i.e., Th and U) are on the order of 10$^{-47}$ m$^2$. Consequently, these particles rarely interact with matter in the earth. The earth's geoneutrino flux is 10$^{25}$ $\APnue$ s$^{-1}$ \cite{mcdonough2020radiogenic}. Each neutrino leaving the earth removes a portion of the earth's radiogenic heat ($Q$).

\noindent \textit{heat producing elements (HPE)}: potassium, thorium, and uranium (i.e., K, Th, and U, or more specifically  $^{40}$K, $^{232}$Th, $^{235}$U, and $^{238}$U) account for $\sim$99.5\% of the earth's radiogenic heating power.

\noindent \textit{Inverse Beta Decay (IBD)}: a nuclear reaction used to detect electron antineutrinos in large underground liquid scintillation detectors that are surrounded by thousands of photomultiplier tubes. The reaction [$\APnue+p\rightarrow e^++n$] involves a free proton (i.e., H atom) capturing a through going $\APnue$ and results in two flashes of light close in space and time. The first flash of light involves $e^+-e^-$ annihilation (order a picosecond following $\APnue+p$ interaction) and the second flash ($\sim$0.2 ms later) comes from the capture by a free proton of the thermalizing neutron. This coincidence of a double light flash in space and time, with the second flash having 2.2 MeV light, reduces the background by a million-fold. This reaction requires the $\APnue$ to carry sufficient energy to overcome the reaction threshold energy of $E_{\bar\nu_e}^{thr}=1.8$~MeV. Thus, restricting us to detecting only antineutrinos from the $\beta^-$ decays in the $^{238}$U and $^{232}$Th decay chains.

\noindent \textit{major component elements}: a cosmochemical classification term for Fe, Ni, Mg, \& Si, with half-mass condensation temperatures ($T_{\mathrm{c}})$ between 1355 and 1250 K. These elements condense from a cooling nebular gas into silicates (first olivine, then pyroxene) and Fe, Ni alloys and together with oxygen represents $\geq$93\% of terrestrial planet's mass \cite{mcdonough2021tp}.

\noindent \textit{Earth and its parts}: the earth is chemically differentiated. It has a metallic core surrounded by the bulk silicate earth (BSE, aka, Primitive Mantle), which initially included the mantle, oceanic and continental crust, and the hydrosphere \& atmosphere; the Primitive Mantle is the undegassed and undifferentiated earth minus the core. The present day silicate earth, less the hydrosphere and atmosphere, is made up of the mantle, including its conductive layers at the bottom (D"), and the top lithosphere; the latter composed of the crust an underlying lithospheric mantle. The lithosphere is the mechanically stiff (i.e., $>$10$^5$ higher viscosity than the underlying asthenospheric mantle), thermally conductive boundary layer. In the continents, the zone above the Moho (a seismically defined boundary between the crust and mantle) is the continental crust and below the continental lithospheric mantle (CLM). Masses and thicknesses of these domains are listed in Table \ref{tab:masses}.

\begin{table}[ht]
\renewcommand\thetable{1}
\caption{Mass of the Earth and its parts }
\label{tab:masses}
\begin{tabular}{lllc}
Domain/reservoir       & thickness & mass                       & citation$^\dagger$ \\
   &  km ($\pm$) &  kg ($\pm$)                       &  \\
\hline
Earth  & 6371 ($^{+7}_{-20}$)   &  5.97218 (60)\;$\times$\;10$^{24}$  & [1] \\
bulk silicate earth (BSE) & 2895 (5)  & 4.036 (6)\;$\times$\;10$^{24}$ & [2]\\
Modern mantle (DM + EM domains)$^*$ & 2867 (20)$^\ddagger$ & 4.002 (20)\;$\times$\;10$^{24}$ & [2] \\
oceanic crust$^\#$    & 10.5 (4.3) & 0.92 (0.11) \;$\times$\;10$^{22}$ & [3] \\
continental crust$^\#$   & 40 (9) & 2.22 (26) \;$\times$\;10$^{22}$ & [3] \\
continental lithospheric mantle (CLM) & 115 (80) & 6.3 (0.8) \;$\times$\;10$^{22}$ & [3]\\
\hline
\end{tabular}
\par\smallskip
$^\dagger$cited source: 1 = \citeA{chambat2010}, 2= \citeA{dziewonski1981preliminary}, 3 = \citeA{Wipperfurth2020}  \\
$^*$DM = Depleted Mantle, the chemically depleted source of MORB (mid-ocean ridge basalts), which is viewed as the chemical complement to the continental crust, and \\ $^*$EM = Enriched Mantle, a smaller (i.e., $\frac{1}{5}$ mass), deeper, and more chemically enriched source of OIB (ocean island basalts).\\
$^\ddagger$from PREM, assuming a uniform surface crust of 24 km \\
$^\#$Using a LITHO1.0 model, see Table 1 in \citeA{Wipperfurth2020}.
\end{table}

\noindent \textit{moderately volatile elements}: a cosmochemical classification term for elements with half-mass condensation temperatures ($T_{\mathrm{c}})$ between 1250 and 600 K. These elements include the alkali metals (lithophile), some transition metals, all the other metals, less Al, and the pnictogens and chalcogens, less N and O.

\noindent \textit{primordial energy}: the energy in the earth from accretion and core formation. Accretion kinetic energy is $\sim$\;10$^{32}$ J, assuming an earth mass (5.97\;$\times$\;10$^{24}$ kg) and 10 km/s as an average velocity of accreting particles. The gravitational energy of core formation, which translates to heating energy, is $\approx$10$^{30}$ J, depending on the assumed settling velocity of a core-forming metal in the growing earth.

\noindent \textit{radiogenic energy}: energy of a nuclear reaction ($Q$) resulting from radioactive decay, given in units of MeV (1 MeV = 10$^6$ eV) or pJ (1 pJ = 10$^{-12}$ J), where 1MeV = 0.1602 pJ. For $\beta$ decays $Q$ (MeV) = (mass$_{p}$ - mass$_{d}$) $\times$\;931.494, with mass$_p$ (mass of parent isotope), mass$_d$ (mass of daughter isotope) in atomic mass units (1 amu = 1.660539\;$\times$\;10$^{-27}$ kg = 931.494 MeV) and for $\alpha$ decay, $Q$ (MeV) = (mass$_{p}$ - mass$_{d}$ - mass$_{\alpha}$)\;$\times$\;931.494, where $E=eV/c^2$, or 1 amu = 0.931494 GeV/c$^2$. 

\noindent \textit{refractory elements}: a cosmochemical classification term for elements with half-mass condensation temperatures ($T_{\mathrm{c}}) >$ 1355 K; they condense at the earliest stage of the cooling of high-temperature nebular gas. These elements are in equal relative proportion ($\pm$15\%) in chondritic meteorites. In terrestrial planets, many of these elements are classified as lithophile (dominantly coupling with oxygen and hosted in the crust and mantle), or siderophile (dominantly metallically bonded and hosted in the core). The refractory elements include: Be, Al, Ca, Ti, Sc, V, Sr, Y, Zr, Nb, Mo, Rh, Ru, Ba, REE, Hf, Ta, W, Re, Os, Ir, Pt, Th and U. The core contains $\geq$90\% of the earth's budget of Mo, Rh, Ru, W, Re, Os, Ir, and Pt, about half of its V, and potentially a minor fraction of its Nb.

\noindent \textit{surface heat flux}: the total surface heat flux from the earth's interior is reported as 46$\pm$3 TW \cite{jaupart:2015tg} or 47$\pm$2 TW \cite{davies2013global}. On average the surface heat flux is about 86 mW/m$^2$, with that for the continents being 65 mW/m$^2$ and for the oceans being 96 mW/m$^2$ \cite{davies2013global}. Energy contributions to this surface flux come from the core (primordial, plus a minor ($\sim$1\% of the surface total) amount due to inner core crystallization), mantle (a combination of primordial and radiogenic), and crust (radiogenic). Other contributions include negligible additions from tidal heating and crust - mantle differentiation.

\section{Detectors and detection technology}

\begin{figure}[h]
    \centering
    \includegraphics[width=0.6\textwidth]{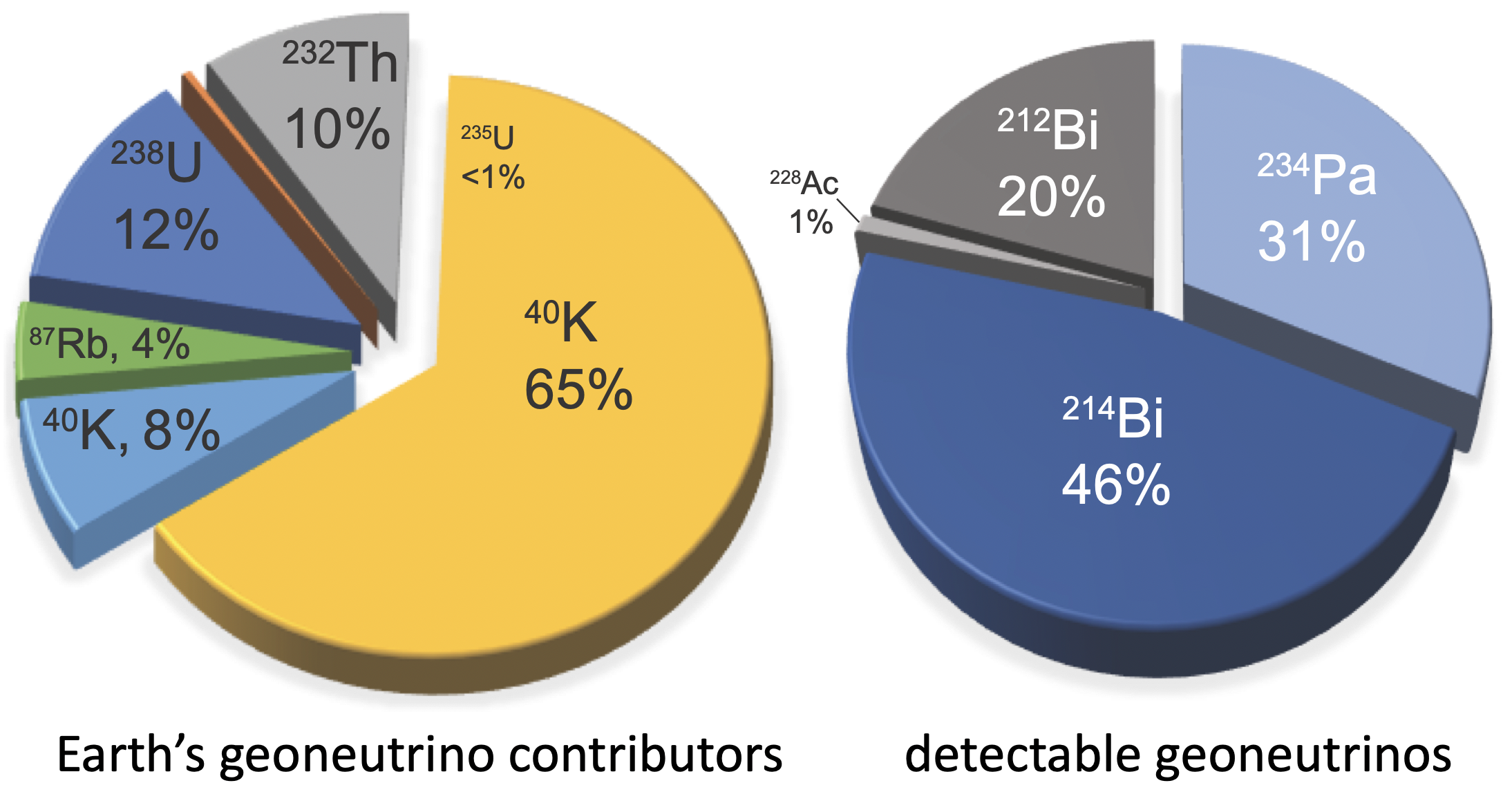}
    \caption{Relative proportions of the earth's present‐day flux of naturally occurring geoneutrinos (left) and the detectable geoneutrinos (right) from the $^{232}$Th decay chain (grey; $^{228}$Ac and $^{212}$Bi) and $^{238}$U decay chain (blue; $^{234}$Pa and $^{214}$Bi, not shown is the negligible contribution from $^{212}$Tl) (see also Table \ref{tab:nu-e}). $^{40}$K has two branches: $\beta^-$ to $^{40}$Ca (65\%) and $EC$ to $^{40}$Ar (8\%).
    }
    \label{fig:contrib}
\end{figure}

Electron antineutrinos ($\APnue$) come mostly from the radioactive decays of $^{40}$K, $^{232}$Th, $^{235}$U, and $^{238}$U \cite<i.e., geoneutrinos; >{KRAUSS}, plus contributions from local anthropogenic sources (i.e., nuclear reactor plants). The earth emits some 10$^{25}$ s$^{-1}$ geoneutrinos \cite{mcdonough2020radiogenic}, with 65\% coming from $\beta^-$ decays of $^{40}$K (Figure \ref{fig:contrib}).

The detection of an electron antineutrino uses the Inverse Beta Decay (IBD) reaction: $\APnue+p\rightarrow e^++n$, which has a reaction threshold energy of $E_{\bar\nu_e}^{thr}=1.806$~MeV.

\begin{equation} \label{threshold}
E_{\bar\nu_e}^{thr}=
\frac{(M_n + m_e)^2 - M^2_p}{2M_p} 
= 1.806[MeV] 
\end{equation}

\noindent assuming the laboratory frame (i.e., stationary target) and where $M_p$, $M_n$, and $m_e$ are the masses of the proton (938.272 MeV), neutron (939.565 MeV) and electron (0.5110 MeV). The neutrino mass is unknown and contributes negligibly to this reaction. Although its upper limit is 0.8 $eV/c^2$ \cite{aker2021direct}, estimates of the neutrino's mass is of the order of 10s to 100s of $meV$ \cite{Nu_mass-ordering18}. This energy threshold restricts the detectable antineutrinos to $\beta^-$ decays from the $^{238}$U and $^{232}$Th decay chains (Figure \ref{fig:contrib}).

\subsection{Technical details for detecting geoneutrinos}
Here we highlight some relevant aspects of the detection scheme.  

Detection occurs when an antineutrino interacts with a free proton, transforming it to a neutron plus a positron, which then causes two flashes of light close in space and time. Each flash of light occurring in these large liquid scintillation spectrometers, which are sited 1 to 2 km underground to shield them from descending, atmospherically produced muons, are detected by the thousands of photomultiplier tubes covering the inner walls, each facing the detector's central volume. 

Energy conservation requires that $E_{\APnue}$ + $M_p$ = $T_{e^+}$ + $m_e$ + $M_n$ + $T_n$, with $T_{e^+}$ and $T_n$ being the kinetic energy of the positron and neutron. The prompt event involves the positron being annihilated in picoseconds by an electron, with the signal being the sum of the reaction releasing a 1.022 MeV energy flash (the sum of the masses of these two leptons) plus the characteristic kinetic energy inherited by the positron from the antineutrino ($E_{prompt}$= ($E_{\APnue}$ + $M_p$ - $m_e$ - $M_n$ - $T_n$) + 2$m_e$, or = $E_{\APnue}$ - $T_n$ - 0.782[MeV]). The accompanying emitted neutron undergoes a cascade of collisions (thermalizing events, loss of energy to its surroundings as it goes towards thermal equilibrium) over about 200 $\mu$s and approximately 15 cm  from the initial interaction point. This neutron is ultimately captured by a second free proton creating $^2$H, resulting in a 2.22 MeV (binding energy) flash. This rare event sequence is eminently detectable because of its characteristics: two flashes of light in space and time, with the second flash having a specific energy. This reaction chain eliminates most background and improves the signal to noise ratio by a million fold.

The organic liquid scintillator is mostly a long chain, aromatic ring hydrocarbon with approximately an H:C proportion of $\sim$2. A wavelength shifting fluor compound is added to the scintillator to set the fluorescence peak emission at $\sim$350-400 nm in order to reach the maximum quantum efficiency of the photomultiplier tubes. The photon yield for the liquid scintillator is typically a light yield of 200 to 400 photons/MeV.

The interaction cross section of antineutrinos (and neutrinos) scales with their energy. For each $\beta^-$ decay there is a spectrum of $\APnue$ emitted energies, which in turn means a spectrum of interaction cross-sections. For IBD detection (i.e., starting at 1.806 MeV), the probability of a geoneutrino detection is low (order 1/10$^{19}$), given that their cross sections are between 10$^{-48}$ and 10$^{-46}$ m$^2$ \cite{vogel1999angular}. The overall emission above the energy threshold level are 0.40 U $\APnue$ per decay and 0.156 Th $\APnue$ per decay. There are four decay chains which produce detectable antineutrinos: two from the $^{232}$Th and two from the $^{238}$U decay chains (Table \ref{tab:nu-e}). The BSE has a Th/U mass ratio of 3.77 (or molar Th/U = 3.90) \cite{wipperfurth:2018}. However, despite Th being four times more abundant than U, attributes of the IBD detection method (i.e., $E_{\APnue}$, branching fraction, and $\sigma_{IBD}$) makes U much easier to detect.

\begin{table}[ht]
\renewcommand\thetable{2}
\begin{threeparttable}
\caption{Detectable $\APnue$ events }
\label{tab:nu-e}
\begin{tabular}{lcccc}
$\beta^-$ decay      & $\APnue$ Max E$^\dagger$ & branching     & Max IBD cross section$^\ddagger$   &    \% of total$^*$\\
events      &   (MeV)  & fraction     &  $\sigma_{IBD}$ (10$^{-43}$ cm$^2$) &    $\APnue$signal\\
\hline
&        &  \textit{$^{232}$Th decay chain}  &  & \\
$^{228}$Ac $\rightarrow$ $^{228}$Th      & 2.134  & 1.00    &  4.3  &   1     \\
$^{212}$Bi $\rightarrow$ $^{212}$Po      & 2.252  & 0.64    &  4.8  &   20     \\
&        & \textit{$^{238}$U decay chain}   &   &  \\
$^{234}$Pa $\rightarrow$ $^{234}$U     & 2.197  & 1.00     & 4.6      &  31     \\
$^{214}$Bi  $\rightarrow$ $^{214}$Po     & 3.270  & 1.00   &    33   &  46    \\
$^{212}$Tl $\rightarrow$ $^{212}$Po      & 4.391  & 0.0002    &  90  &   $<<$1\%     \\
\hline
\end{tabular}
\par\smallskip
$^\dagger$ there is a spectrum of energies for each antineutrino generated during a $\beta^-$ decay, where typically the $\APnue$ takes about $\frac{2}{3}$ and the $e^-$ about $\frac{1}{3}$ of the reaction energy ($Q$). \\
$^\ddagger$as the energy of the antineutrino decreases so does its interaction cross-section.\\
$^*$numbers from Tables 3 and 5 in \citeA{fiorentini2007geo}.
\end{threeparttable}
\end{table}

\subsection{Detectors: Existing, being built, being planned}
\begin{figure}[h]
    \centering
    \includegraphics[width=0.9\textwidth]{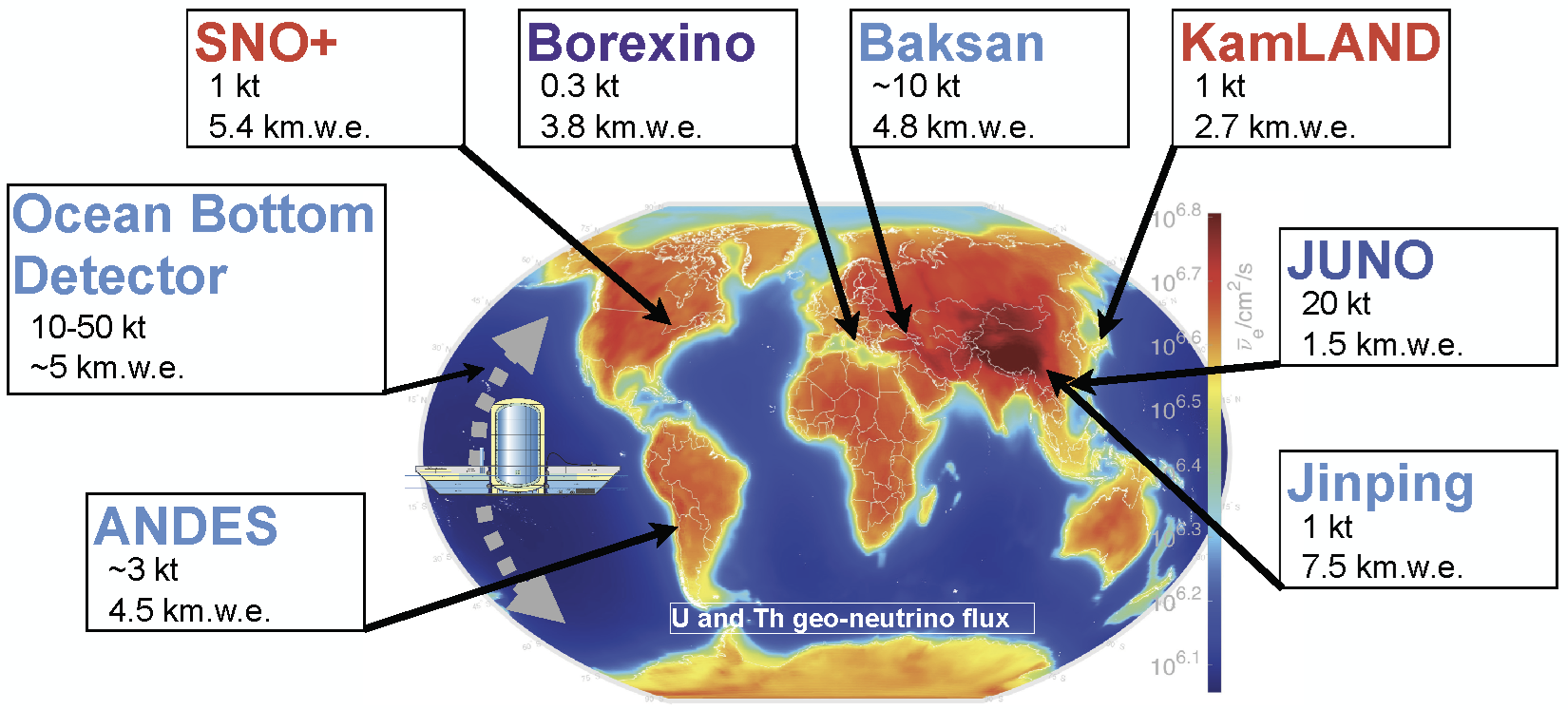}
    \vspace{-0.5cm}
    \caption{Present day global distribution of detectors counting geoneutrino (red), have counted (purple) and are in the development and/or planning stage (blue). JUNO (bold blue) is in the construction phase. Background figure is the calculated global geoneutrino flux (order 10$^6$ $\APnue$ cm$^{-2}$ s$^{-1}$) \cite{Usman2015}; the relatively high flux density seen in the Himalayas is directly correlated with its greater crustal thickness. }
    \label{fig:detectors}
\end{figure}

Currently there are two detectors (Fig. \ref{fig:detectors}) measuring the earth's geoneutrino flux: KamLAND (1 kton), in Kamioka, Japan \cite{araki2005experimental,KAMLAND19}, and SNO+ (1 kton) in Sudbury, Ontario, Canada \cite{SNO+} experiments. The Borexino detector (0.3 kton), in Gran Sasso, Italy \cite{agostini:2020}, has finished. The JUNO (20 kton) experiment in Jiangmen, Guangdong province, China \cite{an2016} is currently being built. The Jinping ($\sim$4 kton) experiment in Jinping Mountains, Sichuan province, China \cite{Beacom_2017} is in development, with prototype detectors onsite testing future detector materials and technologies. Detectors in the proposal stage include Baksan in the Caucasus mountains in Russia \cite{domogatsky2005neutrino}, Andes in Agua Negra tunnels linking the borders of Chile and Argentina \cite{ANDES}, and a proposed ocean bottom detector. Significantly, the Andes detector is the only proposed detector to be sited in the southern hemisphere.

There are ongoing developments for a ocean-going detector. A team of particle physicists from the University of Hawaii put forth a proposal more than 10 years ago for an ocean bottom detector called Hanohano \cite{Learned}. A team of Japanese particle physicists and engineers and earth scientists from JAMSTEC (Japan Agency for Marine-Earth Science and Technology) are currently moving forward with a project to deploy and test a mobile prototype detectors (``Ocean Bottom Detector" (OBD) scale is not yet set, but envisaged to up to 1 ton) off the coast of Japan. A mobile ocean-going detector offers a complementary measurement to land-based experiments. By sitting in the middle of the Pacific Ocean, 3000 km from South America, 3000 km from Australia, and $\sim$3000 km from the core-mantle boundary, such a detector gets a ``mostly-mantle" signal.

\section{Latest results from the physics experiments}

Results from the physics experiments follow counting statistics, with increasing exposure (time spent counting) uncertainties reduce. These experiments are attentive to systematic and statistical uncertainties and addressed these issues in great detail in their publications. 

The measured geoneutrino flux is reported in $\APnue$ cm$^{-2}$ $\mu s^{-1}$ for the KamLAND experiment and in TNU (Terrestrial Neutrino Unit) for the Borexino experiments. \citeA{Mantovani2004} introduced TNU as a way to normalize the differences between detectors. A 1 TNU signal represents the detection of one event in a 1 kiloton liquid scintillation detector over a one year exposure with a 100\% detection efficiency. A 1 kiloton liquid scintillation detector has $\sim$10$^{32}$ free protons (the detection target). Each detector has its own efficiency relative to a 1 kiloton fiducial volume detector, which accounts for the differences in the size of detector, its photomultiplier coverage, its response efficiency of the scintillator, and other factors. The conversion factor between signal in TNU and flux in $\APnue$ cm$^{-2}$ $\mu s^{-1}$ depends on the Th/U ratio and has a value of 0.11 $\APnue$ cm$^{-2}$ $\mu s^{-1}$ TNU$^{-1}$ for earth's Th/U$_{molar}$ of 3.9. 

The most recent results for the KamLAND and Borexino experiments are 32.1$\pm$5.0 \cite{KAMLAND19} and 47.0$^{+8.4}_{-7.7}$ \cite{agostini:2020} TNU, respectively. [The SNO+ detector began counting in 2020 with a partial filled volume (with delays due to the covid-19 pandemic) and is yet to report their data.] The conversion between TNU and TW depends on the geological model assumed for the distribution of Th and U. Using geological models developed for both experiments, the radiogenic heating of the earth ranges from 14 to 25 TW for KamLAND and 19 to 40 TW for Borexino \cite{Wipperfurth2020,mcdonough2020radiogenic}. The combined KamLAND and Borexino results mildly favors an earth model with 20 TW present‐day total power. Development of these and others geological models for each experiment are discussed later.

\section{Compositional models for the Earth}

Earth scientists estimate the planet's radiogenic power within the bounds of 20$\pm$10 TW (terrawatts, 10$^{12}$ watts), with many favoring estimates between 15 and 25 TW. In contrast, estimates from particle physics experiments (including a 19\% contribution from K, not measured) range from 15.3$^{+4.9}_{-4.9}$ TW (KamLAND, Japan) \cite{KAMLAND19} and 38.2$^{+13.6}_{-12.7}$ TW (Borexino, Italy) \cite{agostini:2020}. The latter estimate exceed most predictions by geologists and so one might dismiss such finding from this emerging field of science. The rub here, however, is that this broad range of TW estimates from particle physicists is mostly due to geological uncertainties associated with characterizing the surrounding $\sim$500 km of lithosphere upon which these detectors are sited. These local lithologies contribute $\sim$40$\%$ of the signal seen at a detector \cite{araki2005experimental,Wipperfurth2020}.

Once a geoneutrino flux is measured at a detector a geological model for the immediately surrounding $\sim$500 km of continental crust is applied. This model is coupled with a model for the global lithosphere and mantle contributions (see \citeA{huang2013reference} for method details). Given that a detector's sensitivity scales with $\textit{1/distance$^2$}$, each experiment is most sensitive to the local flux. By taking the total measured signal and subtracting the contributions from the lithosphere (with all of its unknowns), the mantle's contribution to the planetary radiogenic power is identified. Because of its low abundances of Th \& U relative to the continental crust (i.e., ng/g versus $\mu$g/g), the mantle's $\APnue$ contribution can be approximated as relatively homogeneous globally, with mantle heterogeneities in the signal being of the order 10 to 15\% of its total signal \cite{Sramek2013}.

There are three major models that predict the composition of the BSE, and thus the bulk earth. In general, these models agree that the core has a negligible role in hosting the HPE. Although the earth's core is often invoked as a host for  radioactive HPE, mostly to power the geodynamo, there is a little support for such speculations. Recently, \citeA{wipperfurth:2018} show that a maximum of $<$0.5\% of the earth's budget of Th \& U could be hosted in the core. 
Petrologist have identified K-bearing sulfides that might have been extracted into the earth's core. However this evidence does not demonstrate the existence of potassium in the core, it only allows for its possibility. Such plausibility arguments need to be coupled with corroborating paragenetic evidence that is also free of negating geochemical consequences. In all instances, the extraction of a K-bearing phase into the core, is not supported by other geochemical observations (e.g., no evidence for a range of alkali metals and refractory elements (Ca, REE) in the core). Finally, particle physics experiments have focused on the question of a nuclear reactor in the center of, or the surrounding, the earth's core. Conclusions of these studies limit the power of a geological reactor to $<$3.7 TW \cite{Gando2013} and $<$2.4 TW \cite{agostini:2020} at the 95\% confidence limit.

The three major models can be characterized by their relative heat production (H): $low \ H, medium \ H$, and $high \ H$. Heat production models ranges from 10 to 38 TW, assumes Th/U$_{molar}$ = 3.90 and K/U = 14,000, and have relative heat contributions of $\sim$20\% from K, $\sim$40\% from Th, and $\sim$40\% from U. Given these ratios there is a simple multiplier for these model compositions: a 20 TW model has 20 ng/g U (a 10 TW model has 10 ng/g U), 77 ng/g Th and 280 $\mu$g/g K. Both Th and U are refractory lithophile elements, like Al, Ca, Ti and the REE. Also, the earth has been demonstrated multiple times through elemental and isotopic data to be chondritic \cite{McDonough1995,willig2020constraints}. Therefore, if we know the absolute abundance of one of these elements (e.g., U) in the bulk earth, then we know the abundances of all 36 (e.g., assuming 20 ng/g U in the BSE and (Al/U)$_{chondritic}$ = 1.08$\times$10$^6$, then Al$_{BSE}$=21.7 mg/g). 

The $low \ H$ models typically have heat production levels of about 10 TW.  Models like \citeA{javoy2010chemical} and \citeA{faure2020determination} have 11 TW of radiogenic power and their refractory lithophile elements abundances in the BSE are enriched by 1.5 times CI chondrite. This enrichment factor is equivalent to core separation, meaning their bulk earth model has a CI chondrite refractory element composition. Many of the $low \ H$ models were constructed to explain a putative $^{142}$Nd isotope anomaly for the earth \cite{boyet2005142nd}. Other $low \ H \ to \ intermediate \ H$ models for the earth invoke non-chondritic refractory element abundances due to collisional erosion processes, which involved losing a substantial fraction of the early earth's crust \cite{oneill2008collisional,campbell2012evidence,jackson2013major}. Advocates for these models have largely fallen silent, as a simpler explanation for the $^{142}$Nd isotope anomaly became obvious \cite{burkhardt2016nucleosynthetic,bouvier2016primitive}. Finally, a consequence of the $low \ H$ model is that the mantle has very little radiogenic power. Assuming a BSE with about 10 TW of radiogenic power and the continental crust contains 7.1$^{+2.1}_{-1.6}$ TW (Table \ref{tab:HPEinCC}), then the mantle has only $\sim$3 TW of radiogenic power and thus the dominant driver of the major geodynamic processes is primordial energy. 

The $high \ H$ models typically have heat production levels of $\geq$30 TW. Compositional models proposed by \citeA{turcotte2001thorium} and \citeA{agostini:2020} conclude the BSE as having 30 to 38 TW of radiogenic power.  These models have no parallels in the cosmochemistry of meteorites. They predict that the bulk earth is enriched in refractory lithophile elements by a factor of 2.5 to 3.2 $\times$ CI chondrite. A survey of chondritic meteorites documents enrichment factors for refractory elements ranging from 1 to 2.2 $\times$ CI  \cite{alexander2019quantitative_CC,alexander2019quantitative_NC}. In opposition to $low \ H$ models, the $high \ H$ models have radiogenic energy as dominantly driving the earth's geodynamic processes. 

Most $medium \ H$ models have heat production levels of 20$\pm$5 TW \cite{McDonough1995,palme2014cosmochemical}, have enrichments in refractory elements consistent with that seen in chondritic meteorites, and are built from a residuum‐melt relationship between peridotites (mantle rocks) and basalts (partial melts of the mantle). The BSE's composition comes from the least melt‐depleted peridotites. This compositional model applies to the whole mantle and does not envisage any compositional layering (e.g., upper versus lower mantle domains). Moreover, this conclusions is supported by tomographic images of subducting oceanic slabs that penetrate into the lower mantle, documenting mass transfer between the upper and lower mantle and by inference whole mantle convection. In addition, the methodology used also captures into the primitive composition any potential domains that might have been created in the early earth and since been isolated. $Medium \ H$ models have approximately $\frac{2}{3}$ primordial and $\frac{1}{3}$ radiogenic energy driving the earth's major geodynamic processes (i.e., mantle convection, plate tectonics and the geodynamo). Recently, \citeA{yoshizaki2021earth} also showed that the Earth and Mars are equally enriched in refractory elements by 1.9 $\times$ CI, which might reflect the enrichment levels for all of the terrestrial planets and point to a fundamental attribute of inner solar system's building blocks.

\section{The geological predictions}
All of our models depend upon an accurate description of the abundances and distribution of Th and U in the continents.  The role of the geologist in collaboration with neutrino scientists is to determine this spatial distribution of the HPE \cite{huang2013reference}. Although this task is relatively straightforward, it offers many challenges and foremost among them are the difficulties in defining the structure and composition of rocks in the crust beneath the surface, particularly in areas where detectors are sited. Global scale models predicting the composition of the continental crust are available \cite{rudnickCompositionContinentalCrust2014,hackerContinentalLowerCrust2015a,Sammon2021,Wipperfurth2020}, however, differences in these models lead to differences in their geoneutrino signal \cite{Mantovani2004,enomoto2007neutrino,huang2013reference}. 

\citeA{Wipperfurth2020} showed that geophysical characterization of the continental crust, using global seismic models, CRUST 2.0, CRUST 1.0 and LITHO1.0, did not significantly contribute to the uncertainty in models describing the distribution of the heat producing elements in the earth. In contrast, however, the geochemical models were identified as contributing the greatest uncertainty to the model. This finding places the onus of responsibility squarely in the camp of the geochemists and the geologists in terms of understanding the 3D distribution of lithologies and compositions within the crust.

\begin{figure}[h]
    \centering
    \includegraphics[width=0.6\textwidth]{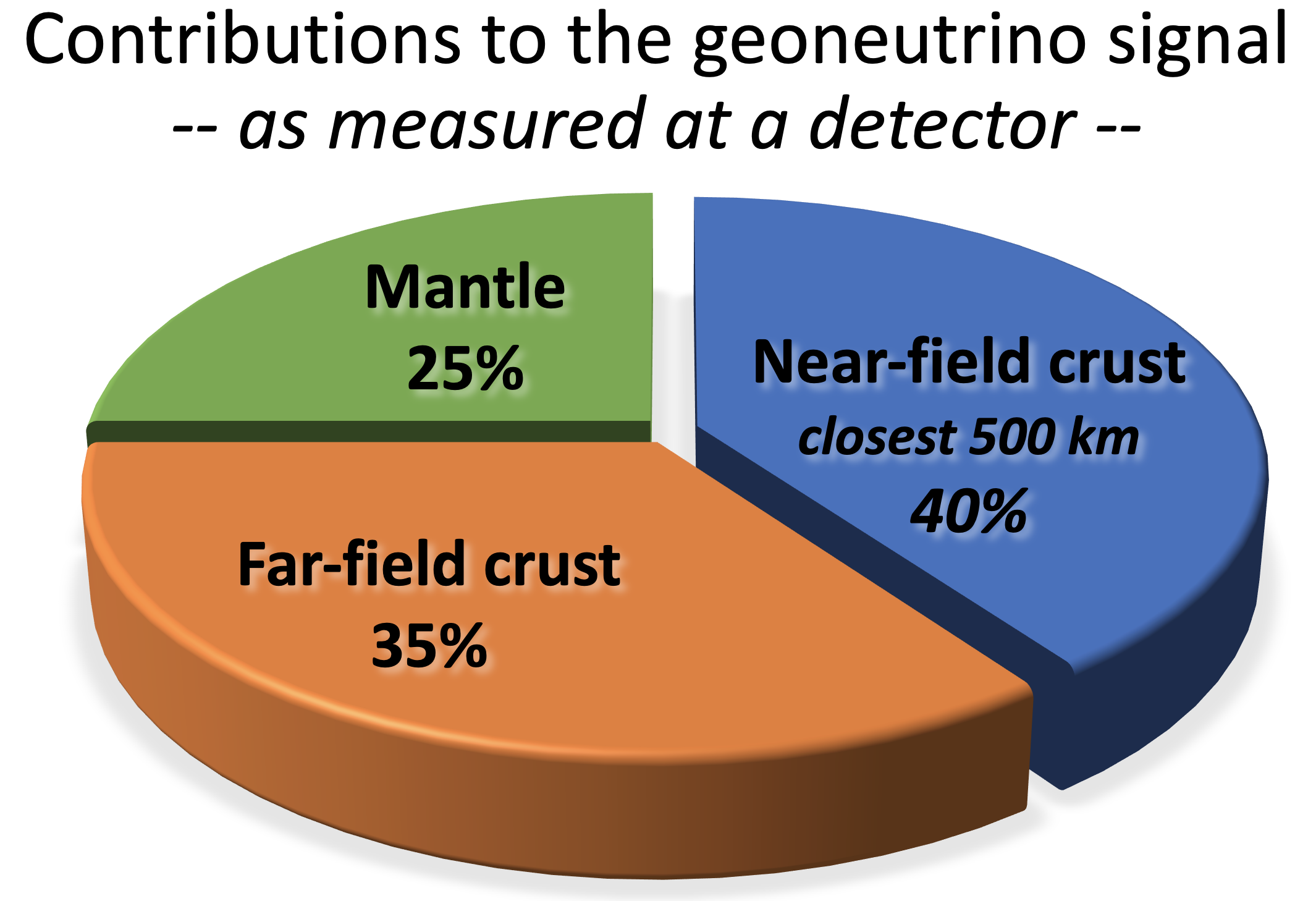}
    \caption{The relative contributions to the total geoneutrino signal: Near-field crust (out to 250 km in all directions), Far-field crust (the rest of the crust around the globe), and Mantle. The local and global Continental Lithospheric Mantle (CLM) contributes $\sim$1 TNU or $\sim$2\% to the total geoneutrino signal at most detector sites. }
    \label{fig:signal-source}
\end{figure}

The total geoneutrino signal represents contributions of three components (Fig. \ref{fig:signal-source}): Near-field crust ($\sim$40\%), Far-field crust ($\sim$35\%, i.e., the global lithospheric signal), and Mantle signals ($\sim$25\%) \cite{Wipperfurth2020}. Paramount in modeling the measured signal at a detector is understanding the contribution from the closest $\sim$500 km of lithosphere adjacent to the detector (i.e., 250 km outwards in any direction from the detector). This region is often referred to as the Near-field, local, or regional crustal signal. This part of the continental crust, and particularly the upper crust, is the region of greatest interest, because of the relative separation distance between source neutrino emitter and the detector, and because it is the brightest geoneutrino emitter due to geological processes that concentrate the HPEs upwards in the continental crust. 

On average the continental crust is enriched in the HPE by more than a factor of 60 over the BSE and $>$100 over the present day abundances in the modern mantle. Moreover, the upper part of the crust is 10 times enriched in the HPE as compared to the lower crust \cite{rudnickCompositionContinentalCrust2014,hackerContinentalLowerCrust2015a,Sammon2021,Wipperfurth2020}. Bulk compositional estimates of the continental crust report that it hosts about 30 to 40\% of the heat producing elements in the BSE (Table \ref{tab:HPEinCC}).


\begin{table}[h]
\renewcommand\thetable{3}
\begin{threeparttable}
\caption{Heat producing elements in the continental crust }
\label{tab:HPEinCC}
\begin{tabular}{r|cccc|c}
     & \multicolumn{4}{l}{  Models of the continental crust}  $|$ & BSE model  \\
\hline 
citation$^\dagger$                & 1           & 2        & 2      & 3           & 4       \\
Estimates of uncertainty          & ($\pm$30\%) & low      & high   & ($\pm$35\%) & ($\pm$10\%) \\
\hline
U ($\mu$g/g) in bulk Cont. crust  & 1.3         & 1.09     & 1.33   & 1.27    & 0.020  \\
Th ($\mu$g/g) in bulk Cont. crust & 5.6         & 4.20     & 5.31   & 5.64    & 0.077 \\
K ($\mu$g/g) in bulk Cont. crust. & 15000       & 11900    & 18800  & 11700   & 280   \\
Heat   production (nW/kg)         & 0.325       & 0.253    & 0.333  & 0.312   & 0.005 \\
Heat production ($\mu$W/m$^3$)$^*$ & 0.943      & 0.733    & 0.967  & 0.906   & 0.023 \\
Radiogenic power (TW)$^\ddagger$  & 7.2         & 5.7      & 7.4    & 7.1     & 20   \\
\% total U in CC                 & 36\%         & 30\%     & 37\%   & 36\%    &   --   \\
\% total Th in CC                & 40\%         & 30\%     & 38\%   & 40\%    &   --   \\
\% total K in CC                 & 29\%         & 23\%     & 37\%   & 23\%    &   --   \\
\hline
\end{tabular}
\par\smallskip
$^\dagger$ 1= \cite{rudnickCompositionContinentalCrust2014}, 2=\cite{hackerContinentalLowerCrust2015a}, 3=\cite{Wipperfurth2020}, and 4=\cite{McDonough1995,Arevalo2009,wipperfurth:2018}. \hspace{0.5cm} $^*$assumes average $\rho$ (density) is 2900 kg/m$^3$ for the continental crust and 4450 kg/m$^3$ for the BSE. 
$^\ddagger$ see Table 1 for reservoir masses.
\end{threeparttable}
\end{table}

The estimated model compositions of the continental crust (Table \ref{tab:HPEinCC}) come with a considerable mount of uncertainty in their predicted abundances. Importantly, all of these models consider the upper crust to represent about the top third of the continent's mass and assume a shared upper crustal compositional model \cite{rudnickCompositionContinentalCrust2014}, which dominates the geoneutrino flux. Using the global geophysical models that describe the continental crust (e.g., LITHO1.0) and a global model for the composition of the continents (Table \ref{tab:HPEinCC}), the geoneutrino signal for the Far-field crust (i.e., the crust minus the local contribution at the detector) ranges from $\sim$8 to 19 TNU, depending on the local setting \cite{Wipperfurth2020}; locations like KamLAND and JUNO, near the coast, fall on the lower end of the scale, where as locations like Jinping and SNO+, surrounded by a significant amount of continental crust, fall on the high end of the range. \citeA{Wipperfurth2020} showed that when combining the data for KamLAND and Borexino and using their geological models, found a mantle signal of 9.2$\pm$8.5 TNU and found the earth's global radiogenic power to be 21.5$\pm$10.4 TW.

The lithospheric signal can be treated globally, but the local lithospheric signal is best done with a detail analysis of the geological, geochemical, and geophysical data of the region.  Using this combination of data one can build a 3D model of the chemical and physical attributes of the local lithosphere. It is noteworthy, however, that such attempts have been conducted at both the KamLAND and Borexino experiment locations. Figure \ref{fig:mantle_flux} and Table \ref{tab:TNU_signal} presents the range of predicted results for the geoneutrino flux at KamLAND, Borexino, and other locations. 

\section{Determining the radioactive power in the mantle}
The particle physics experiment, at its simplest level, reports the geoneutrino flux at a detector. With this number in hand, along with a model for the local and global lithospheric contribution, the physicist then determines the planetary geoneutrino flux and the flux from the mantle. This idea was first laid out by \citeA{KRAUSS}. Later, \citeA{raghavan1998measuring}, in a comparative analyses of signals from detectors in continental and oceanic settings, specifically developed a scheme to determine the mantle flux. 

Surface heat flux observations constrain the earth to radiating 46$\pm$3 TW (i.e., 0.09 W/m$^2$). Based on a compositional model for the continental crust (Table \ref{tab:HPEinCC}), it contains about 7.1$^{+2.1}_{-1.6}$ TW of radiogenic power, leaving $\sim$40 TW as a mantle flux contribution. This latter flux contains contributions from the core (basal heating), which is estimated to be order 10$\pm$5 TW, and mantle (i.e., order 30$\pm$5 TW, which includes radiogenic and primordial additions). A combined analysis of the KamLAND and Borexino results yields a mantle signal of 13$\pm$12 TW \cite{Wipperfurth2020}. Consequently, a large range of compositional models are acceptable for the earth's radiogenic power (e.g., Table \ref{tab:TNU_signal}). 

An important step towards reducing the uncertainties on the estimated mantle radiogenic power would be either (1) determine the mantle flux in oceanic setting and/or (2) improve the accuracy and precision of the local signal (Near-field).  In the next section we address measuring the signal with an ocean bottom detector. Here we consider differences in estimates of the local signal. 

\begin{figure}[h]
    \centering
    \includegraphics[width=0.6\textwidth]{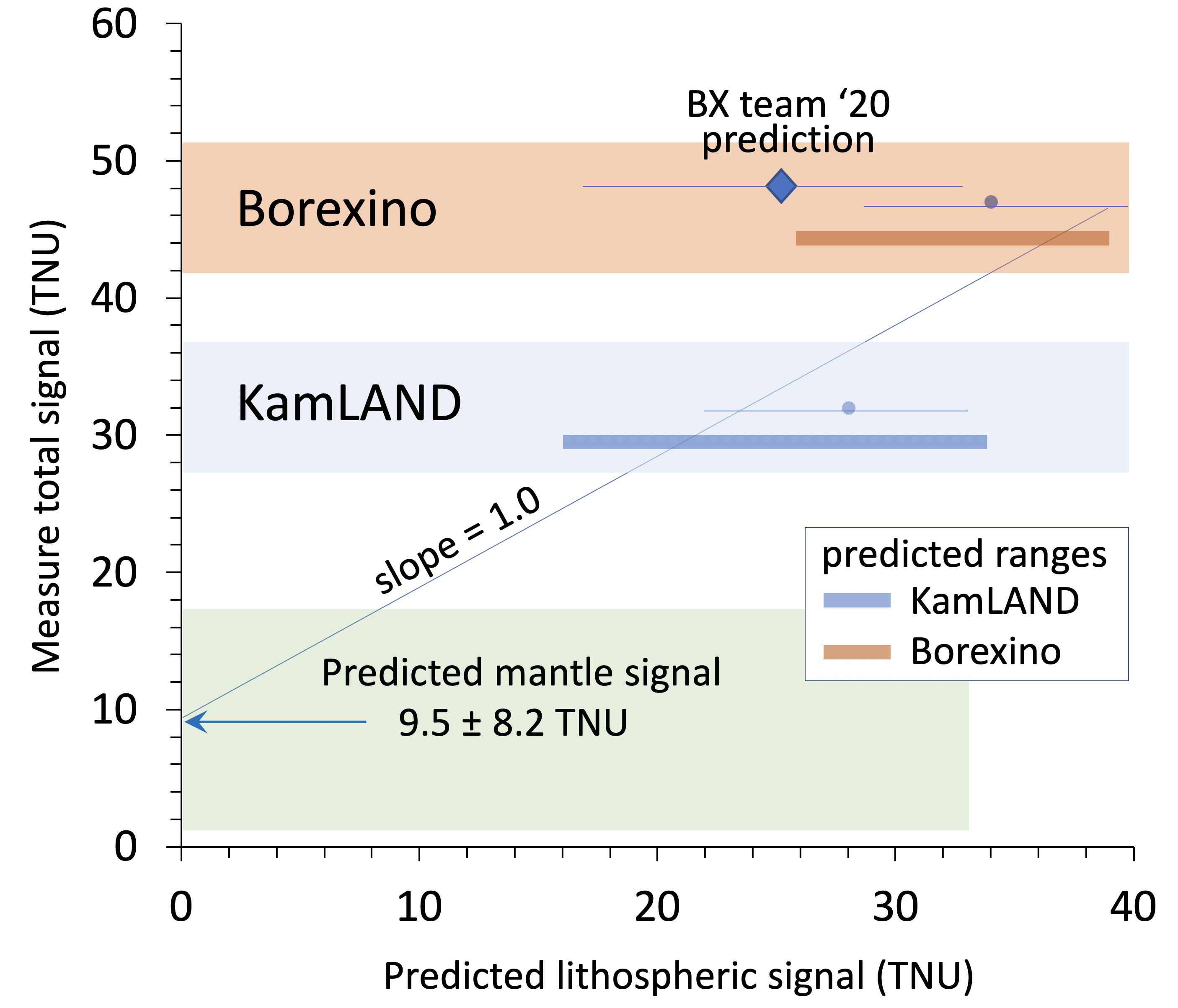}
    \caption{Model predictions for the geoneutrino signal at KamLAND and Borexino and predicted mantle flux (i.e., $\sim$12 TW). See Table \ref{tab:TNU_signal} for data sources and references. KamLAND's measured flux from \citeA{KAMLAND19} (32.1$\pm$5.0 TNU, uncertainty shown as a light blue, horizontal band; y-axis only constraint). Borexino's measured flux from \citeA{agostini:2020} (47.0$^{+8.4}_{-7.7}$ TNU, uncertainty shown as a light orange, horizontal band; y-axis only constraint). The slope=1 is because its a two-component system (crust + mantle). Estimated signals at future detector locations (Table \ref{tab:TNU_signal}) will better constrain the intercept (mantle prediction).
}
    \label{fig:mantle_flux}
\end{figure}

Figure \ref{fig:mantle_flux} and Table \ref{tab:TNU_signal} illustrates the problem. The graph is designed to extract the mantle signal from the combined analysis of data from different detectors.  The signal is made up of crust plus mantle; it is a two component only system (i.e., slope=1). The mantle signal (the ordinate intercept) can be considered, at this scale, homogeneous. Note, \citeA{Sramek2013} found that even when modeling a mantle with 2 large, approximately antipodal structures (e.g., LLSVP: Large low shear velocity provinces), which might contain a factor of 5 or more difference in K, Th, and U abundances, the global mantle signal varied by only $\pm$10\%. Thus, the mantle flux estimate highly depends on the accuracy of the crustal predictions for each detector.

Figure \ref{fig:mantle_flux} also reveals the challenge with different model predictions for the geoneutrino flux from the bulk lithosphere; each model and their uncertainties yields different geoneutrino flux estimates, and in turn this influences the prediction for the mantle signal. It is unclear which model provides a more accurate representation of the crust. In 2020, the Borexino particle physics team reported an updated prediction for the local lithospheric signal and mantle \cite{agostini:2020}, which lead to their distinctly higher prediction (and large uncertainty) for the earth's total radiogenic power (i.e., 38$^{+13.6}_{-12.7}$ TW). Thus, a clear challenge for the geological community is to independently construct a 3D model for the local lithosphere and from this predict a geoneutrino flux. By doing so, it would test this hypothesized lithospheric model constructed by the particle physics team and their markedly radiogenic prediction for the earth's HPE abundances.

\begin{table}[]
\renewcommand\thetable{4}
\begin{threeparttable}
\caption{Geological estimates of the signal contributions at various antineutrino detectors }
\label{tab:TNU_signal}
\begin{tabular}{rlccccc}
\hline
Detector     & reference   & total   & global crustal  & FFC     & NFC    & $\dagger$mantle \\
     &    & signal  & contribution &      &     & contribution \\
\hline \hline 
KamLAND      & \cite{enomoto2007neutrino}  & 38.5             & 28.2          & 10.5     & 17.7      & 10.3 \\
KamLAND      & \cite{huang2013reference}  & 30.7             & 20.6          & 7.3      & 13.3      & 8.8  \\
KamLAND      & \cite{fiorentini2012mantle} & --               & 26.5          & 8.8      & 17.7      & --   \\
KamLAND      & \cite{Wipperfurth2020}     & 37.9             & 27.0          & 8.8      & 18.2      & 9.4  \\
\hline 
Borexino     & \cite{huang2013reference}  & 43.5             & 29.0          & 13.7     & 15.3      & 8.7  \\
Borexino     & \cite{coltorti2011u}       & 43.5             & 26.2          & 16.0     & 10.2      & 9.9  \\
Borexino     & \cite{fiorentini2012mantle} & --               & 25.3          & 15.7     & 9.7       & --   \\
Borexino     & \cite{agostini:2020}       & 47.0             & 25.5          & 16.3     & 9.2       & 20.6 \\
Borexino     & \cite{Wipperfurth2020}    & 43.9             & 32.5          & 14.8     & 18.2      & 9.4  \\
\hline 
SNO+         & \cite{huang2014regional} & 40.0             & 30.7          & 15.1     & 15.6      & 7.0  \\
SNO+         & \cite{strati2017perceiving} & 43.1             & 30.5          & 15.2     & 15.3      & 6.9  \\
SNO+         & \cite{Wipperfurth2020}    & 46.8             & 34.3          & 14.7     & 19.6      & 9.1  \\
\hline 
JUNO         & \cite{strati2015expected} & 39.7             & 28.2          & 13.4     & 17.4      & 8.8  \\
JUNO         & \cite{gao2020juloc}      & 49.1             & 38.3          & 9.8      & 28.5*     & 8.7  \\
JUNO         & \cite{Wipperfurth2020}    & 40.5             & 29.8          & 12.7     & 17.1      & 9.5  \\
\hline 
Jinping      & \cite{sramek2016revealing} & 58.5         & 50.3          & 16.1      & 27.2     & 8.2 \\
Jinping      & \cite{Wan:Jinping}      & 59.4           & 49.0          & --        & --       & 10.4 \\
Jinping      & \cite{Wipperfurth2020}   & 60.0           & 48.8          & 18.7      & 30.3     & 9.3 \\
\hline 
Hanohano     & \cite{huang2013reference} & 12.0          & 2.6           & 2.6       & --       & 9.0 \\
\hline                                                                                          
\end{tabular}
\par\smallskip
All numbers are reported in units of TNU (see text for details). $^\dagger$reported or calculated flux.  NFF = Near Field Crust contribution, FFC = Far Field crust contribution,  *authors defined the JUNO NFC as 10$^{\circ}$ $\times$ 10$^{\circ}$. The flux contribution from the CLM (Continental Lithospheric Mantle) is not included, as it is not always reported by authors; it's contribution is typically on the order of 1 to 2 TNU.
\end{threeparttable}
\end{table}

Future geoneutrino detectors are being built at locations in Guangzhou, China (JUNO) and the eastern slope of the Tibetan plateau (Jinping). Each location has already had a geological model prediction of their geoneutrino signal, which critically depends upon the modeling of the local lithospheric signal (i.e., crust plus lithospheric mantle). The challenge faced by the geological community is how to treat differences in the production of the local lithospheric signal.  Recently, \citeA{strati2015expected} and \citeA{gao2020juloc} reported their predicted both crust signal for the JUNO detector as 28.2$^{+5.2}_{-4.5}$ TNU and 38.3$\pm$4.8 TNU, respectively. These numbers do not overlap with their combined uncertainties and challenge us to resolve this difference. Consequently, the physics experiments are bringing further insights into our understanding of the HPE in the earth, which are forcing geologists to improve their 3D physical and chemical descriptions of the earth.

The question remains -- how to critically assess the accuracy of these competing predictions? Some insights will come by adding more detectors and comparing multiple signals in order to predict a mantle signal. The SNO+ detector has begun counting and so we look forward to adding that data point to Figure \ref{fig:mantle_flux} in the near future. 

\section{Future prospects}

The signal at all of these detectors depends critically upon the signal from the local lithosphere. Given two unknowns parameters (crust and mantle), then the problem is reduced to identifying exclusively the mantle signal. The best way to determine the mantle signal is to detected it far away from continental sources. This means determining the mantle geoneutrino flux from the deep ocean basin likely somewhere in the central Pacific ocean. An ideal location in the central southern Pacific that is approximately a core -- mantle distance ($\sim$3000 km) away from continental masses rimming the Pacific.

The idea of an ocean-going detector that could isolate the mantle signal was introduced  by \citeA{raghavan1998measuring}. In 2007 the Hanohano project \cite{Learned} provided a detailed technical report showing that an ocean-going neutrino detector could be deployed in the deep ocean and serve multiple applications. \citeA{Sramek2013} identified a series of target locations in a north-south transit through the Pacific that could map out potential mantle structures and identifying compositional heterogeneities in the deep mantle.

The concept envisages an ocean-going geoneutrino detector that is deployed on a yearly basis, anchored just above the ocean floor, recovered, serviced, and redeployed. Such a detector would be able to measure the mantle geoneutrino flux over multiple deployments. Given a small lithospheric signal from the Far-field continents and the oceanic crust, it is estimated that 75\% of the signal measured by an ocean bottom detector (OBD) would be from the mantle.

\begin{figure}[h]
    \centering
    \includegraphics[width=0.7\textwidth]{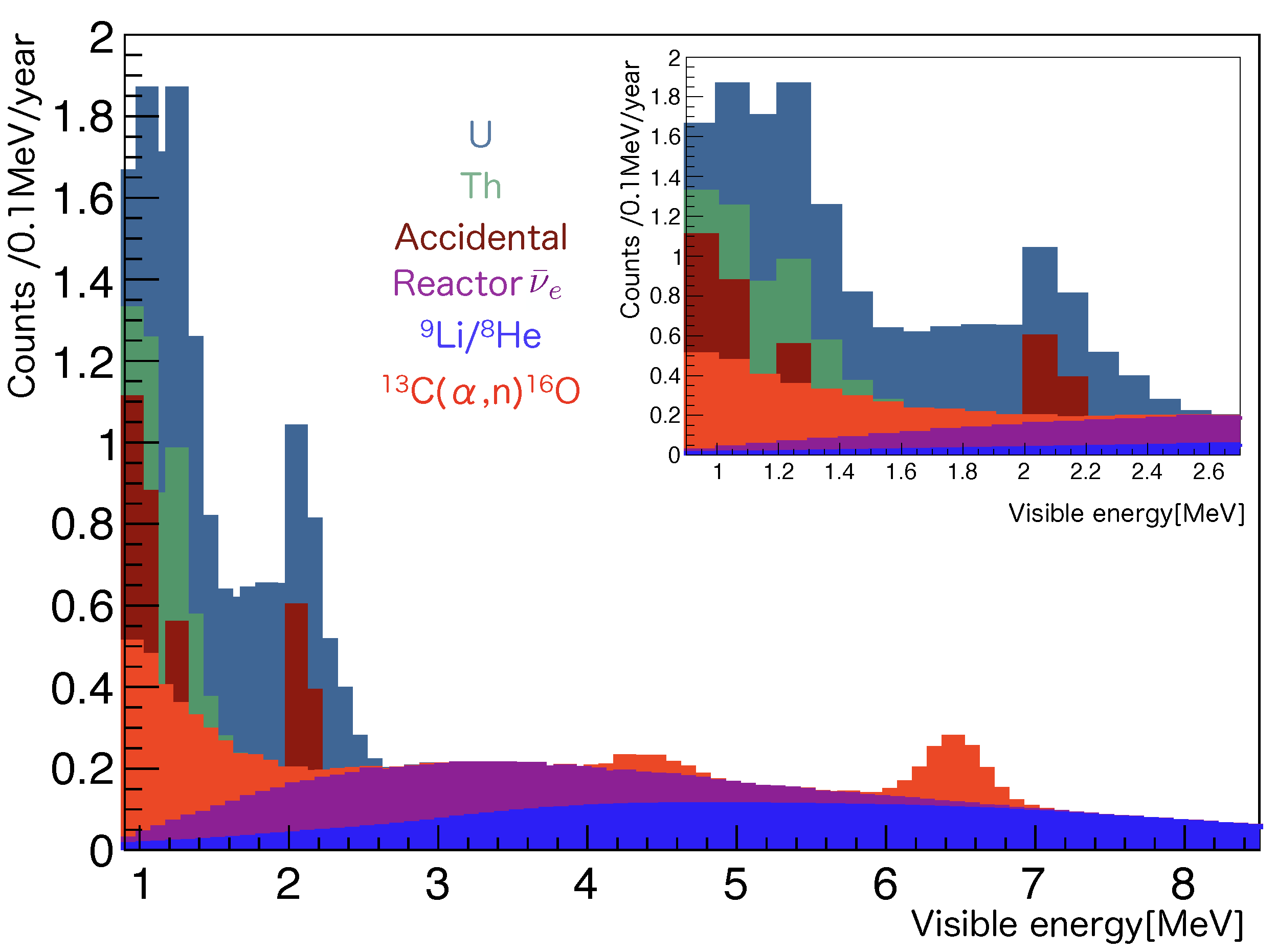}
    \vspace{-0.5cm}
    \caption{Expected energy spectrum of 1.5~kt OBD estimated by detector simulation.  Right top panel focuses on geoneutrino energy range.
}
    \label{fig:OBD_energy}
\end{figure}

Table \ref{tab:OBD_event} and Figure \ref{fig:OBD_energy} summarise estimated signals and backgrounds for 1.5~kt OBD by detector simulation.  The detector was assumed to be deployed off the coast of Hawaii at the depth of 2.7~km.  Radioactive contamination in the detector components, such as liquid scintillator, acrylic vessel, photomultiplier tube, and its water pressure resistant vessel, can be the sources of the backgrounds (e.g. accidental and $^{13}$C($\alpha$,n)$^{16}$O). Detector cleanness controls how much scintillator mass we can use as the fiducial volume.  Seawater acts as a shield for cosmic-ray muons which induce radioactive isotopes (e.g. $^{9}$Li, $^{8}$He). The deeper a detector can be anchored just above the seafloor, the greater the reduction in the muon-related background.  A 1.5~kt OBD has sensitivity to measure mantle geoneutrino at 3.4$\sigma$ level with 3-year exposure measurement. A simple comparison between OBD and [KamLAND \cite{KAMLAND19}], with both as 1.5 kton detectors,  yields a geoneutrino flux of 7 [5] and 9 [24] events/year for mantle and non-mantle (including the crust and other sources listed in Table \ref{tab:OBD_event}) signals, respectively.


Such an instrument is capable of accurately determining the mantle geoneutrino flux, which in turned can be used to unravel the geological signals measured at land-based experiments. Given this, we would have the capability to interpret the local lithospheric signal from all of the existing and future detectors.

\begin{table}[h]
\centering
\caption{Estimated 1-year signals and backgrounds for 1.5 kt OBD by detector simulation. Simulation assumed the detector was deployed at off coast of Hawaii in the depth of 2.7 km. To reduce the radioactive backgrounds from the detector surface, fiducial cut was applied (72~cm).}
\label{tab:OBD_event}
\begin{tabular}{lllll}
\multicolumn{2}{c||}{Signals}                                                 & \multicolumn{3}{c}{Backgrounds}                                                                                                              \\ \hline
\multicolumn{1}{c}{\multirow{2}{*}{Source}} & \multicolumn{1}{c||}{Rate$^\dagger$}      & \multicolumn{1}{c}{\multirow{2}{*}{Source}} & \multicolumn{2}{c}{Rate$^\dagger$}   \\
\multicolumn{1}{c}{}                        & \multicolumn{1}{c||}{(Mantle)}   & \multicolumn{1}{c}{}                        & \textless{}8.5 MeV & \begin{tabular}[c]{@{}l@{}}\textless{}2.6 MeV\\ (geo $\bar{\nu}_{e}$ region)\end{tabular} \\ \hline
U                                           & \multicolumn{1}{l||}{7.4 (5.5)} & Reactor $\bar{\nu}_{e}$                                   & 4.5                & 1.7                                                                       \\
Th                                          & \multicolumn{1}{l||}{1.8 (1.3)} & Accidental                                  & 1.8                & 1.8                                                                       \\ \cline{1-2}
Total                                       & \multicolumn{1}{l||}{9.2 (6.8)} & $^{9}$Li, $^{8}$He                                     & 6.2                & 0.6                                                                       \\
                                            & \multicolumn{1}{l||}{}          & $^{13}$C($\alpha$,n)$^{16}$O                             & 3.6                & 2.6                                                                       \\ \cline{3-5} 
                                            & \multicolumn{1}{l||}{}          & Total                                       & 16.1               & 6.7                                                                      
\end{tabular}
\par\smallskip
$^\dagger$ Rate is in units of counted events/year. \\
Geoneutrino and reactor neutrino spectrum from \cite{Dye2015} 
\end{table}

Neutrino Geoscience compares geoneutrino flux measurements from particle physics experiments with flux predictions derived from integrating data from geology, geophysics, and geochemistry. The particle physics community are focusing on improving their measurement precision and detector technology. Efforts in geosciences are focused on improving our models for the abundance and distribution of the heat producing elements. Improvements in crustal studies are of greatest need. Global and regional geochemical studies, including studies of crustal terrains and xenoliths can help to constrain the composition of the deep crust. 

Holistic approaches are essential for developing improved models of the crust. For example, considerable promise come from studies that incorporate thermodynamic modeling (e.g., Perple\_X; \citeA{connolly2005}) with geochemical and seismic data, in combination with constraints from crustal geotherms (e.g., surface heat flux studies, curie depth maps, reflection data estimates on Moho temperatures, etc) all in an effort to determine crustal compositional trends from the surface to the Moho \cite{sammon2020lower,takeuchi2019stochastic}. The other geological priority needing improvements is  reducing or resolving the up to 30\% spread in crustal predictions of the geoneutrino flux for all but the SNO+ experiment (Table \ref{tab:TNU_signal}).

\section*{Author Contributions}
\noindent Both authors discussed the concepts introduced in the paper. The writing of the main text was principally conducted by WFM; both authors were involved in the calculations and editing process. The authors have read and approved this manuscript.

\section*{Data Availability Statement}
\noindent The data used in this manuscript is freely available upon request.
\acknowledgments
WFM gratefully acknowledges NSF grant support (EAR1650365 and EAR2050374). HW gratefully acknowledges grants (JP15H05833 and JP20H01909) from the Japan Society for the Promotion of Science.  WFM \& HW appreciate the reviewer's comments and are very thankful to the two anonymous reviewer's and the editor handling our paper.
There are no financial or other conflicts of interest with this work.

\bibliography{geonu.bib}

\begin{thebibliography}{}

\bibitem [\protect \citeauthoryear {%
Agostini%
\ \protect \BOthers {.}}{%
Agostini%
\ \protect \BOthers {.}}{%
{\protect \APACyear {2020}}%
}]{%
agostini:2020}
\APACinsertmetastar {%
agostini:2020}%
\begin{APACrefauthors}%
Agostini, M.%
, Altenm{\"u}ller, K.%
, Appel, S.%
, Atroshchenko, V.%
, Bagdasarian, Z.%
, Basilico, D.%
\BDBL {}Zuzel, G.%
\end{APACrefauthors}%
\unskip\
\newblock
\APACrefYearMonthDay{2020}{}{}.
\newblock
{\BBOQ}\APACrefatitle {Comprehensive geoneutrino analysis with {B}orexino}
  {Comprehensive geoneutrino analysis with {B}orexino}.{\BBCQ}
\newblock
\APACjournalVolNumPages{Phys. Rev. D}{101}{1}{012009}.
\newblock
\begin{APACrefDOI} \doi{10.1103/PhysRevD.101.012009} \end{APACrefDOI}
\PrintBackRefs{\CurrentBib}

\bibitem [\protect \citeauthoryear {%
Aker%
\ \protect \BOthers {.}}{%
Aker%
\ \protect \BOthers {.}}{%
{\protect \APACyear {2021}}%
}]{%
aker2021direct}
\APACinsertmetastar {%
aker2021direct}%
\begin{APACrefauthors}%
Aker, M.%
, Beglarian, A.%
, Behrens, J.%
, Berlev, A.%
, Besserer, U.%
, Bieringer, B.%
\BDBL {}Zeller, G.%
\end{APACrefauthors}%
\unskip\
\newblock
\APACrefYearMonthDay{2021}{}{}.
\newblock
{\BBOQ}\APACrefatitle {First direct neutrino-mass measurement with sub-{eV}
  sensitivity} {First direct neutrino-mass measurement with sub-{eV}
  sensitivity}.{\BBCQ}
\newblock
\APACjournalVolNumPages{arXiv preprint}{}{}{arXiv:2105.08533}.
\PrintBackRefs{\CurrentBib}

\bibitem [\protect \citeauthoryear {%
Alexander%
}{%
Alexander%
}{%
{\protect \APACyear {2019}}%
{\protect \APACexlab {{\protect \BCnt {1}}}}}]{%
alexander2019quantitative_CC}
\APACinsertmetastar {%
alexander2019quantitative_CC}%
\begin{APACrefauthors}%
Alexander, C\BPBI M\BPBI O.%
\end{APACrefauthors}%
\unskip\
\newblock
\APACrefYearMonthDay{2019{\protect \BCnt {1}}}{}{}.
\newblock
{\BBOQ}\APACrefatitle {{Quantitative models for the elemental and isotopic
  fractionations in chondrites: The carbonaceous chondrites}} {{Quantitative
  models for the elemental and isotopic fractionations in chondrites: The
  carbonaceous chondrites}}.{\BBCQ}
\newblock
\APACjournalVolNumPages{Geochimica et Cosmochimica Acta}{254}{}{277--309}.
\newblock
\begin{APACrefDOI} \doi{10.1016/j.gca.2019.02.008} \end{APACrefDOI}
\PrintBackRefs{\CurrentBib}

\bibitem [\protect \citeauthoryear {%
Alexander%
}{%
Alexander%
}{%
{\protect \APACyear {2019}}%
{\protect \APACexlab {{\protect \BCnt {2}}}}}]{%
alexander2019quantitative_NC}
\APACinsertmetastar {%
alexander2019quantitative_NC}%
\begin{APACrefauthors}%
Alexander, C\BPBI M\BPBI O.%
\end{APACrefauthors}%
\unskip\
\newblock
\APACrefYearMonthDay{2019{\protect \BCnt {2}}}{}{}.
\newblock
{\BBOQ}\APACrefatitle {{Quantitative models for the elemental and isotopic
  fractionations in the chondrites: The non-carbonaceous chondrites}}
  {{Quantitative models for the elemental and isotopic fractionations in the
  chondrites: The non-carbonaceous chondrites}}.{\BBCQ}
\newblock
\APACjournalVolNumPages{Geochimica et Cosmochimica Acta}{254}{}{246--276}.
\newblock
\begin{APACrefDOI} \doi{10.1016/j.gca.2019.01.026} \end{APACrefDOI}
\PrintBackRefs{\CurrentBib}

\bibitem [\protect \citeauthoryear {%
An%
\ \protect \BOthers {.}}{%
An%
\ \protect \BOthers {.}}{%
{\protect \APACyear {2016}}%
}]{%
an2016}
\APACinsertmetastar {%
an2016}%
\begin{APACrefauthors}%
An, F.%
, An, G.%
, An, Q.%
, Antonelli, V.%
, Baussan, E.%
, Beacom, J.%
\BDBL {}Zou, J.%
\end{APACrefauthors}%
\unskip\
\newblock
\APACrefYearMonthDay{2016}{}{}.
\newblock
{\BBOQ}\APACrefatitle {Neutrino Physics with {{JUNO}}} {Neutrino physics with
  {{JUNO}}}.{\BBCQ}
\newblock
\APACjournalVolNumPages{Journal of Physics G: Nuclear and Particle
  Physics}{43}{3}{030401}.
\newblock
\begin{APACrefDOI} \doi{10.1088/0954-3899/43/3/030401} \end{APACrefDOI}
\PrintBackRefs{\CurrentBib}

\bibitem [\protect \citeauthoryear {%
Andringa%
\ \protect \BOthers {.}}{%
Andringa%
\ \protect \BOthers {.}}{%
{\protect \APACyear {2016}}%
}]{%
SNO+}
\APACinsertmetastar {%
SNO+}%
\begin{APACrefauthors}%
Andringa, S.%
, Arushanova, E.%
, Asahi, S.%
, Askins, M.%
, Auty, D\BPBI J.%
, Back, A\BPBI R.%
\BDBL {}Zuber, K.%
\end{APACrefauthors}%
\unskip\
\newblock
\APACrefYearMonthDay{2016}{}{}.
\newblock
{\BBOQ}\APACrefatitle {Current status and future prospects of the {SNO+}
  experiment} {Current status and future prospects of the {SNO+}
  experiment}.{\BBCQ}
\newblock
\APACjournalVolNumPages{Advances in High Energy Physics}{2016}{}{6194250}.
\newblock
\begin{APACrefDOI} \doi{10.1155/2016/6194250} \end{APACrefDOI}
\PrintBackRefs{\CurrentBib}

\bibitem [\protect \citeauthoryear {%
Araki%
\ \protect \BOthers {.}}{%
Araki%
\ \protect \BOthers {.}}{%
{\protect \APACyear {2005}}%
}]{%
araki2005experimental}
\APACinsertmetastar {%
araki2005experimental}%
\begin{APACrefauthors}%
Araki, T.%
, Enomoto, S.%
, Furuno, K.%
, Gando, Y.%
, Ichimura, K.%
, Ikeda, H.%
\BDBL {}others%
\end{APACrefauthors}%
\unskip\
\newblock
\APACrefYearMonthDay{2005}{}{}.
\newblock
{\BBOQ}\APACrefatitle {Experimental investigation of geologically produced
  antineutrinos with {KamLAND}} {Experimental investigation of geologically
  produced antineutrinos with {KamLAND}}.{\BBCQ}
\newblock
\APACjournalVolNumPages{Nature}{436}{7050}{499--503}.
\newblock
\begin{APACrefDOI} \doi{10.1038/nature03980} \end{APACrefDOI}
\PrintBackRefs{\CurrentBib}

\bibitem [\protect \citeauthoryear {%
Arevalo%
, McDonough%
\BCBL {}\ \BBA {} Luong%
}{%
Arevalo%
\ \protect \BOthers {.}}{%
{\protect \APACyear {2009}}%
}]{%
Arevalo2009}
\APACinsertmetastar {%
Arevalo2009}%
\begin{APACrefauthors}%
Arevalo, R.%
, McDonough, W\BPBI F.%
\BCBL {}\ \BBA {} Luong, M.%
\end{APACrefauthors}%
\unskip\
\newblock
\APACrefYearMonthDay{2009}{}{}.
\newblock
{\BBOQ}\APACrefatitle {{The K/U} ratio of the silicate {E}arth: Insights into
  mantle composition, structure and thermal evolution} {{The K/U} ratio of the
  silicate {E}arth: Insights into mantle composition, structure and thermal
  evolution}.{\BBCQ}
\newblock
\APACjournalVolNumPages{Earth and Planetary Science
  Letters}{278}{3-4}{361--369}.
\newblock
\begin{APACrefDOI} \doi{10.1016/j.epsl.2008.12.023} \end{APACrefDOI}
\PrintBackRefs{\CurrentBib}

\bibitem [\protect \citeauthoryear {%
Beacom%
\ \protect \BOthers {.}}{%
Beacom%
\ \protect \BOthers {.}}{%
{\protect \APACyear {2017}}%
}]{%
Beacom_2017}
\APACinsertmetastar {%
Beacom_2017}%
\begin{APACrefauthors}%
Beacom, J\BPBI F.%
, Chen, S.%
, Cheng, J.%
, Doustimotlagh, S\BPBI N.%
, Gao, Y.%
, Gong, G.%
\BDBL {}Zuber, K.%
\end{APACrefauthors}%
\unskip\
\newblock
\APACrefYearMonthDay{2017}{}{}.
\newblock
{\BBOQ}\APACrefatitle {Physics prospects of the {Jinping} neutrino experiment}
  {Physics prospects of the {Jinping} neutrino experiment}.{\BBCQ}
\newblock
\APACjournalVolNumPages{Chinese Physics C}{41}{2}{023002}.
\newblock
\begin{APACrefDOI} \doi{10.1088/1674-1137/41/2/023002} \end{APACrefDOI}
\PrintBackRefs{\CurrentBib}

\bibitem [\protect \citeauthoryear {%
Bouvier%
\ \BBA {} Boyet%
}{%
Bouvier%
\ \BBA {} Boyet%
}{%
{\protect \APACyear {2016}}%
}]{%
bouvier2016primitive}
\APACinsertmetastar {%
bouvier2016primitive}%
\begin{APACrefauthors}%
Bouvier, A.%
\BCBT {}\ \BBA {} Boyet, M.%
\end{APACrefauthors}%
\unskip\
\newblock
\APACrefYearMonthDay{2016}{}{}.
\newblock
{\BBOQ}\APACrefatitle {{Primitive Solar System materials and Earth share a
  common initial \ce{^{142}Nd} abundance}} {{Primitive Solar System materials
  and Earth share a common initial \ce{^{142}Nd} abundance}}.{\BBCQ}
\newblock
\APACjournalVolNumPages{Nature}{537}{7620}{399--402}.
\newblock
\begin{APACrefDOI} \doi{10.1038/nature19351} \end{APACrefDOI}
\PrintBackRefs{\CurrentBib}

\bibitem [\protect \citeauthoryear {%
Boyet%
\ \BBA {} Carlson%
}{%
Boyet%
\ \BBA {} Carlson%
}{%
{\protect \APACyear {2005}}%
}]{%
boyet2005142nd}
\APACinsertmetastar {%
boyet2005142nd}%
\begin{APACrefauthors}%
Boyet, M.%
\BCBT {}\ \BBA {} Carlson, R\BPBI W.%
\end{APACrefauthors}%
\unskip\
\newblock
\APACrefYearMonthDay{2005}{}{}.
\newblock
{\BBOQ}\APACrefatitle {{\ce{^{142}Nd} evidence for early ($>$ 4.53 Ga) global
  differentiation of the silicate Earth}} {{\ce{^{142}Nd} evidence for early
  ($>$ 4.53 Ga) global differentiation of the silicate Earth}}.{\BBCQ}
\newblock
\APACjournalVolNumPages{Science}{309}{5734}{576--581}.
\newblock
\begin{APACrefDOI} \doi{10.1126/science.1113634} \end{APACrefDOI}
\PrintBackRefs{\CurrentBib}

\bibitem [\protect \citeauthoryear {%
Burkhardt%
\ \protect \BOthers {.}}{%
Burkhardt%
\ \protect \BOthers {.}}{%
{\protect \APACyear {2016}}%
}]{%
burkhardt2016nucleosynthetic}
\APACinsertmetastar {%
burkhardt2016nucleosynthetic}%
\begin{APACrefauthors}%
Burkhardt, C.%
, Borg, L\BPBI E.%
, Brennecka, G\BPBI A.%
, Shollenberger, Q\BPBI R.%
, Dauphas, N.%
\BCBL {}\ \BBA {} Kleine, T.%
\end{APACrefauthors}%
\unskip\
\newblock
\APACrefYearMonthDay{2016}{}{}.
\newblock
{\BBOQ}\APACrefatitle {{A nucleosynthetic origin for the Earth's anomalous
  \ce{^{142}Nd} composition}} {{A nucleosynthetic origin for the Earth's
  anomalous \ce{^{142}Nd} composition}}.{\BBCQ}
\newblock
\APACjournalVolNumPages{Nature}{537}{7620}{394--398}.
\newblock
\begin{APACrefDOI} \doi{10.1038/nature18956} \end{APACrefDOI}
\PrintBackRefs{\CurrentBib}

\bibitem [\protect \citeauthoryear {%
Campbell%
\ \BBA {} O'Neill%
}{%
Campbell%
\ \BBA {} O'Neill%
}{%
{\protect \APACyear {2012}}%
}]{%
campbell2012evidence}
\APACinsertmetastar {%
campbell2012evidence}%
\begin{APACrefauthors}%
Campbell, I\BPBI H.%
\BCBT {}\ \BBA {} O'Neill, H\BPBI S\BPBI C.%
\end{APACrefauthors}%
\unskip\
\newblock
\APACrefYearMonthDay{2012}{}{}.
\newblock
{\BBOQ}\APACrefatitle {{Evidence against a chondritic Earth}} {{Evidence
  against a chondritic Earth}}.{\BBCQ}
\newblock
\APACjournalVolNumPages{Nature}{483}{7391}{553--558}.
\newblock
\begin{APACrefDOI} \doi{10.1038/nature10901} \end{APACrefDOI}
\PrintBackRefs{\CurrentBib}

\bibitem [\protect \citeauthoryear {%
Chambat%
, Ricard%
\BCBL {}\ \BBA {} Valette%
}{%
Chambat%
\ \protect \BOthers {.}}{%
{\protect \APACyear {2010}}%
}]{%
chambat2010}
\APACinsertmetastar {%
chambat2010}%
\begin{APACrefauthors}%
Chambat, F.%
, Ricard, Y.%
\BCBL {}\ \BBA {} Valette, B.%
\end{APACrefauthors}%
\unskip\
\newblock
\APACrefYearMonthDay{2010}{}{}.
\newblock
{\BBOQ}\APACrefatitle {Flattening of the {{Earth}}: Further from Hydrostaticity
  than Previously Estimated: {{Hydrostatic}} Flattening} {Flattening of the
  {{Earth}}: Further from hydrostaticity than previously estimated:
  {{Hydrostatic}} flattening}.{\BBCQ}
\newblock
\APACjournalVolNumPages{Geophysical Journal International}{183}{2}{727-732}.
\newblock
\begin{APACrefDOI} \doi{10.1111/j.1365-246X.2010.04771.x} \end{APACrefDOI}
\PrintBackRefs{\CurrentBib}

\bibitem [\protect \citeauthoryear {%
Coltorti%
\ \protect \BOthers {.}}{%
Coltorti%
\ \protect \BOthers {.}}{%
{\protect \APACyear {2011}}%
}]{%
coltorti2011u}
\APACinsertmetastar {%
coltorti2011u}%
\begin{APACrefauthors}%
Coltorti, M.%
, Boraso, R.%
, Mantovani, F.%
, Morsilli, M.%
, Fiorentini, G.%
, Riva, A.%
\BDBL {}others%
\end{APACrefauthors}%
\unskip\
\newblock
\APACrefYearMonthDay{2011}{}{}.
\newblock
{\BBOQ}\APACrefatitle {U and {Th} content in the {Central Apennines}
  continental crust: A contribution to the determination of the geo-neutrinos
  flux at {LNGS}} {U and {Th} content in the {Central Apennines} continental
  crust: A contribution to the determination of the geo-neutrinos flux at
  {LNGS}}.{\BBCQ}
\newblock
\APACjournalVolNumPages{Geochimica et Cosmochimica Acta}{75}{9}{2271--2294}.
\newblock
\begin{APACrefDOI} \doi{10.1016/j.gca.2011.01.024} \end{APACrefDOI}
\PrintBackRefs{\CurrentBib}

\bibitem [\protect \citeauthoryear {%
Connolly%
}{%
Connolly%
}{%
{\protect \APACyear {2005}}%
}]{%
connolly2005}
\APACinsertmetastar {%
connolly2005}%
\begin{APACrefauthors}%
Connolly, J\BPBI A\BPBI D.%
\end{APACrefauthors}%
\unskip\
\newblock
\APACrefYearMonthDay{2005}{}{}.
\newblock
{\BBOQ}\APACrefatitle {Computation of Phase Equilibria by Linear Programming:
  {{A}} Tool for Geodynamic Modeling and Its Application to Subduction Zone
  Decarbonation} {Computation of phase equilibria by linear programming: {{A}}
  tool for geodynamic modeling and its application to subduction zone
  decarbonation}.{\BBCQ}
\newblock
\APACjournalVolNumPages{Earth and Planetary Science Letters}{236}{1}{524-541}.
\newblock
\begin{APACrefDOI} \doi{10.1016/j.epsl.2005.04.033} \end{APACrefDOI}
\PrintBackRefs{\CurrentBib}

\bibitem [\protect \citeauthoryear {%
Davies%
}{%
Davies%
}{%
{\protect \APACyear {2013}}%
}]{%
davies2013global}
\APACinsertmetastar {%
davies2013global}%
\begin{APACrefauthors}%
Davies, J\BPBI H.%
\end{APACrefauthors}%
\unskip\
\newblock
\APACrefYearMonthDay{2013}{}{}.
\newblock
{\BBOQ}\APACrefatitle {Global map of solid Earth surface heat flow} {Global map
  of solid earth surface heat flow}.{\BBCQ}
\newblock
\APACjournalVolNumPages{Geochemistry, Geophysics,
  Geosystems}{14}{10}{4608--4622}.
\newblock
\begin{APACrefDOI} \doi{10.1002/ggge.20271} \end{APACrefDOI}
\PrintBackRefs{\CurrentBib}

\bibitem [\protect \citeauthoryear {%
de Salas%
, Gariazzo%
, Mena%
, Ternes%
\BCBL {}\ \BBA {} Tórtola%
}{%
de Salas%
\ \protect \BOthers {.}}{%
{\protect \APACyear {2018}}%
}]{%
Nu_mass-ordering18}
\APACinsertmetastar {%
Nu_mass-ordering18}%
\begin{APACrefauthors}%
de Salas, P\BPBI F.%
, Gariazzo, S.%
, Mena, O.%
, Ternes, C\BPBI A.%
\BCBL {}\ \BBA {} Tórtola, M.%
\end{APACrefauthors}%
\unskip\
\newblock
\APACrefYearMonthDay{2018}{}{}.
\newblock
{\BBOQ}\APACrefatitle {Neutrino Mass Ordering from Oscillations and Beyond:
  2018 Status and Future Prospects} {Neutrino mass ordering from oscillations
  and beyond: 2018 status and future prospects}.{\BBCQ}
\newblock
\APACjournalVolNumPages{Frontiers in Astronomy and Space Sciences}{5}{}{36}.
\newblock
\begin{APACrefDOI} \doi{10.3389/fspas.2018.00036} \end{APACrefDOI}
\PrintBackRefs{\CurrentBib}

\bibitem [\protect \citeauthoryear {%
Dib%
}{%
Dib%
}{%
{\protect \APACyear {2015}}%
}]{%
ANDES}
\APACinsertmetastar {%
ANDES}%
\begin{APACrefauthors}%
Dib, C\BPBI O.%
\end{APACrefauthors}%
\unskip\
\newblock
\APACrefYearMonthDay{2015}{}{}.
\newblock
{\BBOQ}\APACrefatitle {{ANDES}: An Underground Laboratory in {S}outh {A}merica}
  {{ANDES}: An underground laboratory in {S}outh {A}merica}.{\BBCQ}
\newblock
\APACjournalVolNumPages{Physics Procedia}{61}{}{534-541}.
\newblock
\begin{APACrefDOI} \doi{10.1016/j.phpro.2014.12.118} \end{APACrefDOI}
\PrintBackRefs{\CurrentBib}

\bibitem [\protect \citeauthoryear {%
Domogatsky%
, Kopeikin%
, Mikaelyan%
\BCBL {}\ \BBA {} Sinev%
}{%
Domogatsky%
\ \protect \BOthers {.}}{%
{\protect \APACyear {2005}}%
}]{%
domogatsky2005neutrino}
\APACinsertmetastar {%
domogatsky2005neutrino}%
\begin{APACrefauthors}%
Domogatsky, G.%
, Kopeikin, V.%
, Mikaelyan, L.%
\BCBL {}\ \BBA {} Sinev, V.%
\end{APACrefauthors}%
\unskip\
\newblock
\APACrefYearMonthDay{2005}{}{}.
\newblock
{\BBOQ}\APACrefatitle {Neutrino geophysics at {Baksan I}: possible detection of
  georeactor antineutrinos} {Neutrino geophysics at {Baksan I}: possible
  detection of georeactor antineutrinos}.{\BBCQ}
\newblock
\APACjournalVolNumPages{Physics of Atomic Nuclei}{68}{1}{69--72}.
\newblock
\begin{APACrefDOI} \doi{10.1134/1.1858559} \end{APACrefDOI}
\PrintBackRefs{\CurrentBib}

\bibitem [\protect \citeauthoryear {%
Dye%
\ \BBA {} Barna%
}{%
Dye%
\ \BBA {} Barna%
}{%
{\protect \APACyear {2015}}%
}]{%
Dye2015}
\APACinsertmetastar {%
Dye2015}%
\begin{APACrefauthors}%
Dye, S.%
\BCBT {}\ \BBA {} Barna, A.%
\end{APACrefauthors}%
\unskip\
\newblock
\APACrefYearMonthDay{2015}{}{}.
\newblock
\APACrefbtitle {Global Antineutrino Modeling for a Web Application.} {Global
  antineutrino modeling for a web application.}
\newblock
\APACrefnote{arXiv:1510.05633}
\PrintBackRefs{\CurrentBib}

\bibitem [\protect \citeauthoryear {%
Dziewonski%
\ \BBA {} Anderson%
}{%
Dziewonski%
\ \BBA {} Anderson%
}{%
{\protect \APACyear {1981}}%
}]{%
dziewonski1981preliminary}
\APACinsertmetastar {%
dziewonski1981preliminary}%
\begin{APACrefauthors}%
Dziewonski, A\BPBI M.%
\BCBT {}\ \BBA {} Anderson, D\BPBI L.%
\end{APACrefauthors}%
\unskip\
\newblock
\APACrefYearMonthDay{1981}{}{}.
\newblock
{\BBOQ}\APACrefatitle {{Preliminary reference Earth model}} {{Preliminary
  reference Earth model}}.{\BBCQ}
\newblock
\APACjournalVolNumPages{Physics of the Earth and Planetary
  Interiors}{25}{4}{297--356}.
\newblock
\begin{APACrefDOI} \doi{10.1016/0031-9201(81)90046-7} \end{APACrefDOI}
\PrintBackRefs{\CurrentBib}

\bibitem [\protect \citeauthoryear {%
Enomoto%
, Ohtani%
, Inoue%
\BCBL {}\ \BBA {} Suzuki%
}{%
Enomoto%
\ \protect \BOthers {.}}{%
{\protect \APACyear {2007}}%
}]{%
enomoto2007neutrino}
\APACinsertmetastar {%
enomoto2007neutrino}%
\begin{APACrefauthors}%
Enomoto, S.%
, Ohtani, E.%
, Inoue, K.%
\BCBL {}\ \BBA {} Suzuki, A.%
\end{APACrefauthors}%
\unskip\
\newblock
\APACrefYearMonthDay{2007}{}{}.
\newblock
{\BBOQ}\APACrefatitle {Neutrino geophysics with {KamLAND} and future prospects}
  {Neutrino geophysics with {KamLAND} and future prospects}.{\BBCQ}
\newblock
\APACjournalVolNumPages{Earth and Planetary Science
  Letters}{258}{1-2}{147--159}.
\newblock
\begin{APACrefDOI} \doi{10.1016/j.epsl.2007.03.038} \end{APACrefDOI}
\PrintBackRefs{\CurrentBib}

\bibitem [\protect \citeauthoryear {%
Faure%
\ \protect \BOthers {.}}{%
Faure%
\ \protect \BOthers {.}}{%
{\protect \APACyear {2020}}%
}]{%
faure2020determination}
\APACinsertmetastar {%
faure2020determination}%
\begin{APACrefauthors}%
Faure, P.%
, Boyet, M.%
, Bouhifd, M.%
, Manthilake, G.%
, Hammouda, T.%
\BCBL {}\ \BBA {} Devidal, J\BHBI L.%
\end{APACrefauthors}%
\unskip\
\newblock
\APACrefYearMonthDay{2020}{}{}.
\newblock
{\BBOQ}\APACrefatitle {{Determination of the refractory enrichment factor of
  the bulk silicate Earth from metal-silicate experiments on rare earth
  elements}} {{Determination of the refractory enrichment factor of the bulk
  silicate Earth from metal-silicate experiments on rare earth
  elements}}.{\BBCQ}
\newblock
\APACjournalVolNumPages{Earth and Planetary Science Letters}{}{}{116644}.
\newblock
\begin{APACrefDOI} \doi{10.1016/j.epsl.2020.116644} \end{APACrefDOI}
\PrintBackRefs{\CurrentBib}

\bibitem [\protect \citeauthoryear {%
Fiorentini%
, Fogli%
, Lisi%
, Mantovani%
\BCBL {}\ \BBA {} Rotunno%
}{%
Fiorentini%
\ \protect \BOthers {.}}{%
{\protect \APACyear {2012}}%
}]{%
fiorentini2012mantle}
\APACinsertmetastar {%
fiorentini2012mantle}%
\begin{APACrefauthors}%
Fiorentini, G.%
, Fogli, G.%
, Lisi, E.%
, Mantovani, F.%
\BCBL {}\ \BBA {} Rotunno, A.%
\end{APACrefauthors}%
\unskip\
\newblock
\APACrefYearMonthDay{2012}{}{}.
\newblock
{\BBOQ}\APACrefatitle {Mantle geoneutrinos in {KamLAND and Borexino}} {Mantle
  geoneutrinos in {KamLAND and Borexino}}.{\BBCQ}
\newblock
\APACjournalVolNumPages{Physical Review D}{86}{3}{033004}.
\newblock
\begin{APACrefDOI} \doi{10.1103/PhysRevD.86.033004} \end{APACrefDOI}
\PrintBackRefs{\CurrentBib}

\bibitem [\protect \citeauthoryear {%
Fiorentini%
, Lissia%
\BCBL {}\ \BBA {} Mantovani%
}{%
Fiorentini%
\ \protect \BOthers {.}}{%
{\protect \APACyear {2007}}%
}]{%
fiorentini2007geo}
\APACinsertmetastar {%
fiorentini2007geo}%
\begin{APACrefauthors}%
Fiorentini, G.%
, Lissia, M.%
\BCBL {}\ \BBA {} Mantovani, F.%
\end{APACrefauthors}%
\unskip\
\newblock
\APACrefYearMonthDay{2007}{}{}.
\newblock
{\BBOQ}\APACrefatitle {Geo-neutrinos and earth's interior} {Geo-neutrinos and
  earth's interior}.{\BBCQ}
\newblock
\APACjournalVolNumPages{Physics Reports}{453}{5-6}{117--172}.
\newblock
\begin{APACrefDOI} \doi{10.1016/j.physrep.2007.09.001} \end{APACrefDOI}
\PrintBackRefs{\CurrentBib}

\bibitem [\protect \citeauthoryear {%
Gando%
\ \protect \BOthers {.}}{%
Gando%
\ \protect \BOthers {.}}{%
{\protect \APACyear {2013}}%
}]{%
Gando2013}
\APACinsertmetastar {%
Gando2013}%
\begin{APACrefauthors}%
Gando, A.%
, Gando, Y.%
, Hanakago, H.%
, Ikeda, H.%
, Inoue, K.%
, Ishidoshiro, K.%
\BDBL {}Decowski, M\BPBI P.%
\end{APACrefauthors}%
\unskip\
\newblock
\APACrefYearMonthDay{2013}{}{}.
\newblock
{\BBOQ}\APACrefatitle {Reactor on-off antineutrino measurement with {KamLAND}}
  {Reactor on-off antineutrino measurement with {KamLAND}}.{\BBCQ}
\newblock
\APACjournalVolNumPages{Phys. Rev. D}{88}{}{033001}.
\newblock
\begin{APACrefDOI} \doi{10.1103/PhysRevD.88.033001} \end{APACrefDOI}
\PrintBackRefs{\CurrentBib}

\bibitem [\protect \citeauthoryear {%
Gao%
\ \protect \BOthers {.}}{%
Gao%
\ \protect \BOthers {.}}{%
{\protect \APACyear {2020}}%
}]{%
gao2020juloc}
\APACinsertmetastar {%
gao2020juloc}%
\begin{APACrefauthors}%
Gao, R.%
, Li, Z.%
, Han, R.%
, Wang, A.%
, Li, Y.%
, Xi, Y.%
\BDBL {}Xu, Y.%
\end{APACrefauthors}%
\unskip\
\newblock
\APACrefYearMonthDay{2020}{}{}.
\newblock
{\BBOQ}\APACrefatitle {{JULOC}: A local 3-{D} high-resolution crustal model in
  {South China} for forecasting geoneutrino measurements at {JUNO}} {{JULOC}: A
  local 3-{D} high-resolution crustal model in {South China} for forecasting
  geoneutrino measurements at {JUNO}}.{\BBCQ}
\newblock
\APACjournalVolNumPages{Physics of the Earth and Planetary
  Interiors}{299}{}{106409}.
\newblock
\begin{APACrefDOI} \doi{10.1016/j.pepi.2019.106409} \end{APACrefDOI}
\PrintBackRefs{\CurrentBib}

\bibitem [\protect \citeauthoryear {%
Hacker%
, Kelemen%
\BCBL {}\ \BBA {} Behn%
}{%
Hacker%
\ \protect \BOthers {.}}{%
{\protect \APACyear {2015}}%
}]{%
hackerContinentalLowerCrust2015a}
\APACinsertmetastar {%
hackerContinentalLowerCrust2015a}%
\begin{APACrefauthors}%
Hacker, B\BPBI R.%
, Kelemen, P\BPBI B.%
\BCBL {}\ \BBA {} Behn, M\BPBI D.%
\end{APACrefauthors}%
\unskip\
\newblock
\APACrefYearMonthDay{2015}{}{}.
\newblock
{\BBOQ}\APACrefatitle {Continental Lower Crust} {Continental lower
  crust}.{\BBCQ}
\newblock
\APACjournalVolNumPages{Annual Review of Earth and Planetary
  Sciences}{43}{1}{167-205}.
\newblock
\begin{APACrefDOI} \doi{10.1146/annurev-earth-050212-124117} \end{APACrefDOI}
\PrintBackRefs{\CurrentBib}

\bibitem [\protect \citeauthoryear {%
Huang%
, Chubakov%
, Mantovani%
, Rudnick%
\BCBL {}\ \BBA {} McDonough%
}{%
Huang%
\ \protect \BOthers {.}}{%
{\protect \APACyear {2013}}%
}]{%
huang2013reference}
\APACinsertmetastar {%
huang2013reference}%
\begin{APACrefauthors}%
Huang, Y.%
, Chubakov, V.%
, Mantovani, F.%
, Rudnick, R\BPBI L.%
\BCBL {}\ \BBA {} McDonough, W\BPBI F.%
\end{APACrefauthors}%
\unskip\
\newblock
\APACrefYearMonthDay{2013}{}{}.
\newblock
{\BBOQ}\APACrefatitle {A reference {E}arth model for the heat-producing
  elements and associated geoneutrino flux} {A reference {E}arth model for the
  heat-producing elements and associated geoneutrino flux}.{\BBCQ}
\newblock
\APACjournalVolNumPages{Geochemistry, Geophysics,
  Geosystems}{14}{6}{2003--2029}.
\newblock
\begin{APACrefDOI} \doi{10.1002/ggge.20129} \end{APACrefDOI}
\PrintBackRefs{\CurrentBib}

\bibitem [\protect \citeauthoryear {%
Huang%
, Strati%
, Mantovani%
, Shirey%
\BCBL {}\ \BBA {} McDonough%
}{%
Huang%
\ \protect \BOthers {.}}{%
{\protect \APACyear {2014}}%
}]{%
huang2014regional}
\APACinsertmetastar {%
huang2014regional}%
\begin{APACrefauthors}%
Huang, Y.%
, Strati, V.%
, Mantovani, F.%
, Shirey, S\BPBI B.%
\BCBL {}\ \BBA {} McDonough, W\BPBI F.%
\end{APACrefauthors}%
\unskip\
\newblock
\APACrefYearMonthDay{2014}{}{}.
\newblock
{\BBOQ}\APACrefatitle {Regional study of the {Archean to Proterozoic} crust at
  the {Sudbury Neutrino Observatory (SNO+)}, {O}ntario: {P}redicting the
  geoneutrino flux} {Regional study of the {Archean to Proterozoic} crust at
  the {Sudbury Neutrino Observatory (SNO+)}, {O}ntario: {P}redicting the
  geoneutrino flux}.{\BBCQ}
\newblock
\APACjournalVolNumPages{Geochemistry, Geophysics,
  Geosystems}{15}{10}{3925--3944}.
\newblock
\begin{APACrefDOI} \doi{10.1002/2014GC005397} \end{APACrefDOI}
\PrintBackRefs{\CurrentBib}

\bibitem [\protect \citeauthoryear {%
Jackson%
\ \BBA {} Jellinek%
}{%
Jackson%
\ \BBA {} Jellinek%
}{%
{\protect \APACyear {2013}}%
}]{%
jackson2013major}
\APACinsertmetastar {%
jackson2013major}%
\begin{APACrefauthors}%
Jackson, M\BPBI G.%
\BCBT {}\ \BBA {} Jellinek, A\BPBI M.%
\end{APACrefauthors}%
\unskip\
\newblock
\APACrefYearMonthDay{2013}{}{}.
\newblock
{\BBOQ}\APACrefatitle {Major and trace element composition of the high
  {$^3$He/$^4$He} mantle: {Implications} for the composition of a nonchonditic
  Earth} {Major and trace element composition of the high {$^3$He/$^4$He}
  mantle: {Implications} for the composition of a nonchonditic earth}.{\BBCQ}
\newblock
\APACjournalVolNumPages{Geochemistry, Geophysics,
  Geosystems}{14}{8}{2954--2976}.
\newblock
\begin{APACrefDOI} \doi{10.1002/ggge.20188} \end{APACrefDOI}
\PrintBackRefs{\CurrentBib}

\bibitem [\protect \citeauthoryear {%
Jaupart%
, Labrosse%
, Lucazeau%
\BCBL {}\ \BBA {} Mareschal%
}{%
Jaupart%
\ \protect \BOthers {.}}{%
{\protect \APACyear {2015}}%
}]{%
jaupart:2015tg}
\APACinsertmetastar {%
jaupart:2015tg}%
\begin{APACrefauthors}%
Jaupart, C.%
, Labrosse, S.%
, Lucazeau, F.%
\BCBL {}\ \BBA {} Mareschal, J\BHBI C.%
\end{APACrefauthors}%
\unskip\
\newblock
\APACrefYearMonthDay{2015}{}{}.
\newblock
{\BBOQ}\APACrefatitle {Temperatures, Heat, and Energy in the Mantle of the
  {E}arth} {Temperatures, heat, and energy in the mantle of the
  {E}arth}.{\BBCQ}
\newblock
\BIn{} D.~Bercovici\ (\BED), \APACrefbtitle {Mantle Dynamics} {Mantle
  dynamics}\ (\BVOL~7, \BPG~223-270).
\newblock
\APACaddressPublisher{Oxford}{Elsevier}.
\newblock
\APACrefnote{Editor-in-chief G. Schubert}
\newblock
\begin{APACrefDOI} \doi{10.1016/B978-0-444-53802-4.00126-3} \end{APACrefDOI}
\PrintBackRefs{\CurrentBib}

\bibitem [\protect \citeauthoryear {%
Javoy%
\ \protect \BOthers {.}}{%
Javoy%
\ \protect \BOthers {.}}{%
{\protect \APACyear {2010}}%
}]{%
javoy2010chemical}
\APACinsertmetastar {%
javoy2010chemical}%
\begin{APACrefauthors}%
Javoy, M.%
, Kaminski, E.%
, Guyot, F.%
, Andrault, D.%
, Sanloup, C.%
, Moreira, M.%
\BDBL {}Jaupart, C.%
\end{APACrefauthors}%
\unskip\
\newblock
\APACrefYearMonthDay{2010}{}{}.
\newblock
{\BBOQ}\APACrefatitle {{The chemical composition of the Earth: Enstatite
  chondrite models}} {{The chemical composition of the Earth: Enstatite
  chondrite models}}.{\BBCQ}
\newblock
\APACjournalVolNumPages{Earth and Planetary Science Letters}{293}{3}{259--268}.
\newblock
\begin{APACrefDOI} \doi{10.1016/j.epsl.2010.02.033} \end{APACrefDOI}
\PrintBackRefs{\CurrentBib}

\bibitem [\protect \citeauthoryear {%
Krauss%
, Glashow%
\BCBL {}\ \BBA {} Schramm%
}{%
Krauss%
\ \protect \BOthers {.}}{%
{\protect \APACyear {1984}}%
}]{%
KRAUSS}
\APACinsertmetastar {%
KRAUSS}%
\begin{APACrefauthors}%
Krauss, L\BPBI M.%
, Glashow, S\BPBI L.%
\BCBL {}\ \BBA {} Schramm, D\BPBI N.%
\end{APACrefauthors}%
\unskip\
\newblock
\APACrefYearMonthDay{1984}{}{}.
\newblock
{\BBOQ}\APACrefatitle {Anti-neutrinos Astronomy and Geophysics} {Anti-neutrinos
  astronomy and geophysics}.{\BBCQ}
\newblock
\APACjournalVolNumPages{Nature}{310}{}{191-198}.
\newblock
\begin{APACrefDOI} \doi{10.1038/310191a0} \end{APACrefDOI}
\PrintBackRefs{\CurrentBib}

\bibitem [\protect \citeauthoryear {%
Learned%
, Dye%
\BCBL {}\ \BBA {} Pakvasa%
}{%
Learned%
\ \protect \BOthers {.}}{%
{\protect \APACyear {2007}}%
}]{%
Learned}
\APACinsertmetastar {%
Learned}%
\begin{APACrefauthors}%
Learned, J\BPBI G.%
, Dye, S\BPBI T.%
\BCBL {}\ \BBA {} Pakvasa, S.%
\end{APACrefauthors}%
\unskip\
\newblock
\APACrefYearMonthDay{2007}{}{}.
\newblock
{\BBOQ}\APACrefatitle {Hanohano: A Deep ocean anti-neutrino detector for unique
  neutrino physics and geophysics studies} {Hanohano: A deep ocean
  anti-neutrino detector for unique neutrino physics and geophysics
  studies}.{\BBCQ}
\newblock
\BIn{} \APACrefbtitle {Neutrino telescopes. {P}roceedings, 12$^{th}$
  {I}nternational {W}orkshop, {V}enice, {I}taly, {M}arch 6-9, 2007} {Neutrino
  telescopes. {P}roceedings, 12$^{th}$ {I}nternational {W}orkshop, {V}enice,
  {I}taly, {M}arch 6-9, 2007}\ (\BPG~235-269).
\newblock
\APACrefnote{arXiv:0810.4975}
\PrintBackRefs{\CurrentBib}

\bibitem [\protect \citeauthoryear {%
Mantovani%
, Carmignani%
, Fiorentini%
\BCBL {}\ \BBA {} Lissia%
}{%
Mantovani%
\ \protect \BOthers {.}}{%
{\protect \APACyear {2004}}%
}]{%
Mantovani2004}
\APACinsertmetastar {%
Mantovani2004}%
\begin{APACrefauthors}%
Mantovani, F.%
, Carmignani, L.%
, Fiorentini, G.%
\BCBL {}\ \BBA {} Lissia, M.%
\end{APACrefauthors}%
\unskip\
\newblock
\APACrefYearMonthDay{2004}{}{}.
\newblock
{\BBOQ}\APACrefatitle {Antineutrinos from {E}arth: A reference model and its
  uncertainties} {Antineutrinos from {E}arth: A reference model and its
  uncertainties}.{\BBCQ}
\newblock
\APACjournalVolNumPages{Phys. Rev. D}{69}{1}{013001}.
\newblock
\begin{APACrefDOI} \doi{10.1103/PhysRevD.69.013001} \end{APACrefDOI}
\PrintBackRefs{\CurrentBib}

\bibitem [\protect \citeauthoryear {%
McDonough%
\ \BBA {} Sun%
}{%
McDonough%
\ \BBA {} Sun%
}{%
{\protect \APACyear {1995}}%
}]{%
McDonough1995}
\APACinsertmetastar {%
McDonough1995}%
\begin{APACrefauthors}%
McDonough, W\BPBI F.%
\BCBT {}\ \BBA {} Sun, S\BPBI S.%
\end{APACrefauthors}%
\unskip\
\newblock
\APACrefYearMonthDay{1995}{}{}.
\newblock
{\BBOQ}\APACrefatitle {The Composition of the {E}arth} {The composition of the
  {E}arth}.{\BBCQ}
\newblock
\APACjournalVolNumPages{Chemical Geology}{120}{3-4}{223--253}.
\newblock
\begin{APACrefDOI} \doi{10.1016/0009-2541(94)00140-4} \end{APACrefDOI}
\PrintBackRefs{\CurrentBib}

\bibitem [\protect \citeauthoryear {%
{McDonough}%
, {{\v{S}}r{\'a}mek}%
\BCBL {}\ \BBA {} {Wipperfurth}%
}{%
{McDonough}%
\ \protect \BOthers {.}}{%
{\protect \APACyear {2020}}%
}]{%
mcdonough2020radiogenic}
\APACinsertmetastar {%
mcdonough2020radiogenic}%
\begin{APACrefauthors}%
{McDonough}, W\BPBI F.%
, {{\v{S}}r{\'a}mek}, O.%
\BCBL {}\ \BBA {} {Wipperfurth}, S\BPBI A.%
\end{APACrefauthors}%
\unskip\
\newblock
\APACrefYearMonthDay{2020}{}{}.
\newblock
{\BBOQ}\APACrefatitle {{Radiogenic power and geoneutrino luminosity of the
  Earth and other terrestrial bodies through time}} {{Radiogenic power and
  geoneutrino luminosity of the Earth and other terrestrial bodies through
  time}}.{\BBCQ}
\newblock
\APACjournalVolNumPages{Geochemistry, Geophysics,
  Geosystems}{21}{7}{e2019GC008865}.
\newblock
\begin{APACrefDOI} \doi{10.1029/2019GC008865} \end{APACrefDOI}
\PrintBackRefs{\CurrentBib}

\bibitem [\protect \citeauthoryear {%
McDonough%
\ \BBA {} Yoshizaki%
}{%
McDonough%
\ \BBA {} Yoshizaki%
}{%
{\protect \APACyear {2021}}%
}]{%
mcdonough2021tp}
\APACinsertmetastar {%
mcdonough2021tp}%
\begin{APACrefauthors}%
McDonough, W\BPBI F.%
\BCBT {}\ \BBA {} Yoshizaki, T.%
\end{APACrefauthors}%
\unskip\
\newblock
\APACrefYearMonthDay{2021}{}{}.
\newblock
{\BBOQ}\APACrefatitle {{Terrestrial planet compositions controlled by accretion
  disk magnetic field}} {{Terrestrial planet compositions controlled by
  accretion disk magnetic field}}.{\BBCQ}
\newblock
\APACjournalVolNumPages{Progress in Earth and Planetary Science}{8}{}{39}.
\newblock
\begin{APACrefDOI} \doi{10.1186/s40645-021-00429-4} \end{APACrefDOI}
\PrintBackRefs{\CurrentBib}

\bibitem [\protect \citeauthoryear {%
O'Neill%
\ \BBA {} Palme%
}{%
O'Neill%
\ \BBA {} Palme%
}{%
{\protect \APACyear {2008}}%
}]{%
oneill2008collisional}
\APACinsertmetastar {%
oneill2008collisional}%
\begin{APACrefauthors}%
O'Neill, H\BPBI S\BPBI C.%
\BCBT {}\ \BBA {} Palme, H.%
\end{APACrefauthors}%
\unskip\
\newblock
\APACrefYearMonthDay{2008}{}{}.
\newblock
{\BBOQ}\APACrefatitle {{Collisional erosion and the non-chondritic composition
  of the terrestrial planets}} {{Collisional erosion and the non-chondritic
  composition of the terrestrial planets}}.{\BBCQ}
\newblock
\APACjournalVolNumPages{Philosophical Transactions of the Royal Society of
  London A: Mathematical, Physical and Engineering
  Sciences}{366}{1883}{4205--4238}.
\newblock
\begin{APACrefDOI} \doi{10.1098/rsta.2008.0111} \end{APACrefDOI}
\PrintBackRefs{\CurrentBib}

\bibitem [\protect \citeauthoryear {%
Palme%
\ \BBA {} O'Neill%
}{%
Palme%
\ \BBA {} O'Neill%
}{%
{\protect \APACyear {2014}}%
}]{%
palme2014cosmochemical}
\APACinsertmetastar {%
palme2014cosmochemical}%
\begin{APACrefauthors}%
Palme, H.%
\BCBT {}\ \BBA {} O'Neill, H\BPBI S\BPBI C.%
\end{APACrefauthors}%
\unskip\
\newblock
\APACrefYearMonthDay{2014}{}{}.
\newblock
{\BBOQ}\APACrefatitle {Cosmochemical estimates of mantle composition}
  {Cosmochemical estimates of mantle composition}.{\BBCQ}
\newblock
\BIn{} R\BPBI W.~Carlson\ (\BED), \APACrefbtitle {The {Mantle and Core}} {The
  {Mantle and Core}}\ (\BVOL\ 3 of {\em Treatise on Geochemistry (Second
  Edition)}, \BPGS\ 1--39).
\newblock
\APACaddressPublisher{Oxford}{Elsevier}.
\newblock
\APACrefnote{Editors-in-chief H. D. Holland and K. K. Turekian}
\newblock
\begin{APACrefDOI} \doi{10.1016/B978-0-08-095975-7.00201-1} \end{APACrefDOI}
\PrintBackRefs{\CurrentBib}

\bibitem [\protect \citeauthoryear {%
Raghavan%
\ \protect \BOthers {.}}{%
Raghavan%
\ \protect \BOthers {.}}{%
{\protect \APACyear {1998}}%
}]{%
raghavan1998measuring}
\APACinsertmetastar {%
raghavan1998measuring}%
\begin{APACrefauthors}%
Raghavan, R\BPBI S.%
, Schoenert, S.%
, Enomoto, S.%
, Shirai, J.%
, Suekane, F.%
\BCBL {}\ \BBA {} Suzuki, A.%
\end{APACrefauthors}%
\unskip\
\newblock
\APACrefYearMonthDay{1998}{}{}.
\newblock
{\BBOQ}\APACrefatitle {Measuring the global radioactivity in the {Earth} by
  multidetector antineutrino spectroscopy} {Measuring the global radioactivity
  in the {Earth} by multidetector antineutrino spectroscopy}.{\BBCQ}
\newblock
\APACjournalVolNumPages{Physical review letters}{80}{3}{635}.
\newblock
\begin{APACrefDOI} \doi{10.1103/PhysRevLett.80.635} \end{APACrefDOI}
\PrintBackRefs{\CurrentBib}

\bibitem [\protect \citeauthoryear {%
Rudnick%
\ \BBA {} Gao%
}{%
Rudnick%
\ \BBA {} Gao%
}{%
{\protect \APACyear {2014}}%
}]{%
rudnickCompositionContinentalCrust2014}
\APACinsertmetastar {%
rudnickCompositionContinentalCrust2014}%
\begin{APACrefauthors}%
Rudnick, R\BPBI L.%
\BCBT {}\ \BBA {} Gao, S.%
\end{APACrefauthors}%
\unskip\
\newblock
\APACrefYearMonthDay{2014}{}{}.
\newblock
{\BBOQ}\APACrefatitle {Composition of the Continental Crust} {Composition of
  the continental crust}.{\BBCQ}
\newblock
\BIn{} R.~Rudnick\ (\BED), \APACrefbtitle {The {Crust}} {The {Crust}}\ (\BVOL\
  4 of {\em Treatise on Geochemistry (Second Edition)}, \BPG~1-51).
\newblock
\APACaddressPublisher{}{{Elsevier}}.
\newblock
\APACrefnote{Editors-in-chief H. D. Holland and K. K. Turekian}
\newblock
\begin{APACrefDOI} \doi{10.1016/B978-0-08-095975-7.00301-6} \end{APACrefDOI}
\PrintBackRefs{\CurrentBib}

\bibitem [\protect \citeauthoryear {%
L.~Sammon%
, Gao%
\BCBL {}\ \BBA {} McDonough%
}{%
L.~Sammon%
\ \protect \BOthers {.}}{%
{\protect \APACyear {2020}}%
}]{%
sammon2020lower}
\APACinsertmetastar {%
sammon2020lower}%
\begin{APACrefauthors}%
Sammon, L.%
, Gao, C.%
\BCBL {}\ \BBA {} McDonough, W.%
\end{APACrefauthors}%
\unskip\
\newblock
\APACrefYearMonthDay{2020}{}{}.
\newblock
{\BBOQ}\APACrefatitle {Lower crustal composition in the southwestern {United
  States}} {Lower crustal composition in the southwestern {United
  States}}.{\BBCQ}
\newblock
\APACjournalVolNumPages{Journal of Geophysical Research: Solid
  Earth}{125}{3}{e2019JB019011}.
\newblock
\begin{APACrefDOI} \doi{10.1029/2019JB019011} \end{APACrefDOI}
\PrintBackRefs{\CurrentBib}

\bibitem [\protect \citeauthoryear {%
L\BPBI G.~Sammon%
\ \BBA {} McDonough%
}{%
L\BPBI G.~Sammon%
\ \BBA {} McDonough%
}{%
{\protect \APACyear {2021}}%
}]{%
Sammon2021}
\APACinsertmetastar {%
Sammon2021}%
\begin{APACrefauthors}%
Sammon, L\BPBI G.%
\BCBT {}\ \BBA {} McDonough, W\BPBI F.%
\end{APACrefauthors}%
\unskip\
\newblock
\APACrefYearMonthDay{2021}{}{}.
\newblock
{\BBOQ}\APACrefatitle {A Geochemical Review of Amphibolite, Granulite, and
  Eclogite Facies Lithologies: Perspectives on the Deep Continental Crust} {A
  geochemical review of amphibolite, granulite, and eclogite facies
  lithologies: Perspectives on the deep continental crust}.{\BBCQ}
\newblock
\APACjournalVolNumPages{Earth and Space Science Open Archive}{}{}{37}.
\newblock
\begin{APACrefDOI} \doi{10.1002/essoar.10506164.1} \end{APACrefDOI}
\PrintBackRefs{\CurrentBib}

\bibitem [\protect \citeauthoryear {%
{\v S}r{\'a}mek%
\ \protect \BOthers {.}}{%
{\v S}r{\'a}mek%
\ \protect \BOthers {.}}{%
{\protect \APACyear {2013}}%
}]{%
Sramek2013}
\APACinsertmetastar {%
Sramek2013}%
\begin{APACrefauthors}%
{\v S}r{\'a}mek, O.%
, McDonough, W\BPBI F.%
, Kite, E\BPBI S.%
, Leki{\'c}, V.%
, Dye, S\BPBI T.%
\BCBL {}\ \BBA {} Zhong, S.%
\end{APACrefauthors}%
\unskip\
\newblock
\APACrefYearMonthDay{2013}{}{}.
\newblock
{\BBOQ}\APACrefatitle {Geophysical and geochemical constraints on geoneutrino
  fluxes from {E}arth's mantle} {Geophysical and geochemical constraints on
  geoneutrino fluxes from {E}arth's mantle}.{\BBCQ}
\newblock
\APACjournalVolNumPages{Earth and Planetary Science Letters}{361}{}{356--366}.
\newblock
\begin{APACrefDOI} \doi{10.1016/j.epsl.2012.11.001} \end{APACrefDOI}
\PrintBackRefs{\CurrentBib}

\bibitem [\protect \citeauthoryear {%
{\v{S}}r{\'a}mek%
, Roskovec%
, Wipperfurth%
, Xi%
\BCBL {}\ \BBA {} McDonough%
}{%
{\v{S}}r{\'a}mek%
\ \protect \BOthers {.}}{%
{\protect \APACyear {2016}}%
}]{%
sramek2016revealing}
\APACinsertmetastar {%
sramek2016revealing}%
\begin{APACrefauthors}%
{\v{S}}r{\'a}mek, O.%
, Roskovec, B.%
, Wipperfurth, S\BPBI A.%
, Xi, Y.%
\BCBL {}\ \BBA {} McDonough, W\BPBI F.%
\end{APACrefauthors}%
\unskip\
\newblock
\APACrefYearMonthDay{2016}{}{}.
\newblock
{\BBOQ}\APACrefatitle {Revealing the {E}arth’s mantle from the tallest
  mountains using the {J}inping {N}eutrino {E}xperiment} {Revealing the
  {E}arth’s mantle from the tallest mountains using the {J}inping {N}eutrino
  {E}xperiment}.{\BBCQ}
\newblock
\APACjournalVolNumPages{Scientific Reports}{6}{1}{33034}.
\newblock
\begin{APACrefDOI} \doi{10.1038/srep33034} \end{APACrefDOI}
\PrintBackRefs{\CurrentBib}

\bibitem [\protect \citeauthoryear {%
Strati%
\ \protect \BOthers {.}}{%
Strati%
\ \protect \BOthers {.}}{%
{\protect \APACyear {2015}}%
}]{%
strati2015expected}
\APACinsertmetastar {%
strati2015expected}%
\begin{APACrefauthors}%
Strati, V.%
, Baldoncini, M.%
, Callegari, I.%
, Mantovani, F.%
, McDonough, W\BPBI F.%
, Ricci, B.%
\BCBL {}\ \BBA {} Xhixha, G.%
\end{APACrefauthors}%
\unskip\
\newblock
\APACrefYearMonthDay{2015}{}{}.
\newblock
{\BBOQ}\APACrefatitle {Expected geoneutrino signal at {JUNO}} {Expected
  geoneutrino signal at {JUNO}}.{\BBCQ}
\newblock
\APACjournalVolNumPages{Progress in Earth and Planetary Science}{2}{1}{1--7}.
\newblock
\begin{APACrefDOI} \doi{10.1186/s40645-015-0037-6} \end{APACrefDOI}
\PrintBackRefs{\CurrentBib}

\bibitem [\protect \citeauthoryear {%
Strati%
, Wipperfurth%
, Baldoncini%
, McDonough%
\BCBL {}\ \BBA {} Mantovani%
}{%
Strati%
\ \protect \BOthers {.}}{%
{\protect \APACyear {2017}}%
}]{%
strati2017perceiving}
\APACinsertmetastar {%
strati2017perceiving}%
\begin{APACrefauthors}%
Strati, V.%
, Wipperfurth, S\BPBI A.%
, Baldoncini, M.%
, McDonough, W\BPBI F.%
\BCBL {}\ \BBA {} Mantovani, F.%
\end{APACrefauthors}%
\unskip\
\newblock
\APACrefYearMonthDay{2017}{}{}.
\newblock
{\BBOQ}\APACrefatitle {Perceiving the crust in {3-D}: A model integrating
  geological, geochemical, and geophysical data} {Perceiving the crust in
  {3-D}: A model integrating geological, geochemical, and geophysical
  data}.{\BBCQ}
\newblock
\APACjournalVolNumPages{Geochemistry, Geophysics,
  Geosystems}{18}{12}{4326--4341}.
\newblock
\begin{APACrefDOI} \doi{10.1002/2017GC007067} \end{APACrefDOI}
\PrintBackRefs{\CurrentBib}

\bibitem [\protect \citeauthoryear {%
Takeuchi%
\ \protect \BOthers {.}}{%
Takeuchi%
\ \protect \BOthers {.}}{%
{\protect \APACyear {2019}}%
}]{%
takeuchi2019stochastic}
\APACinsertmetastar {%
takeuchi2019stochastic}%
\begin{APACrefauthors}%
Takeuchi, N.%
, Ueki, K.%
, Iizuka, T.%
, Nagao, J.%
, Tanaka, A.%
, Enomoto, S.%
\BDBL {}Tanaka, H\BPBI K.%
\end{APACrefauthors}%
\unskip\
\newblock
\APACrefYearMonthDay{2019}{}{}.
\newblock
{\BBOQ}\APACrefatitle {Stochastic modeling of 3-{D} compositional distribution
  in the crust with {B}ayesian inference and application to geoneutrino
  observation in {J}apan} {Stochastic modeling of 3-{D} compositional
  distribution in the crust with {B}ayesian inference and application to
  geoneutrino observation in {J}apan}.{\BBCQ}
\newblock
\APACjournalVolNumPages{Physics of the Earth and Planetary
  Interiors}{288}{}{37--57}.
\newblock
\begin{APACrefDOI} \doi{10.1016/j.pepi.2019.01.002} \end{APACrefDOI}
\PrintBackRefs{\CurrentBib}

\bibitem [\protect \citeauthoryear {%
Turcotte%
, Paul%
\BCBL {}\ \BBA {} White%
}{%
Turcotte%
\ \protect \BOthers {.}}{%
{\protect \APACyear {2001}}%
}]{%
turcotte2001thorium}
\APACinsertmetastar {%
turcotte2001thorium}%
\begin{APACrefauthors}%
Turcotte, D\BPBI L.%
, Paul, D.%
\BCBL {}\ \BBA {} White, W\BPBI M.%
\end{APACrefauthors}%
\unskip\
\newblock
\APACrefYearMonthDay{2001}{}{}.
\newblock
{\BBOQ}\APACrefatitle {{Thorium-uranium systematics require layered mantle
  convection}} {{Thorium-uranium systematics require layered mantle
  convection}}.{\BBCQ}
\newblock
\APACjournalVolNumPages{Journal of Geophysical Research: Solid
  Earth}{106}{B3}{4265--4276}.
\newblock
\begin{APACrefDOI} \doi{10.1029/2000JB900409} \end{APACrefDOI}
\PrintBackRefs{\CurrentBib}

\bibitem [\protect \citeauthoryear {%
Usman%
, Jocher%
, Dye%
, McDonough%
\BCBL {}\ \BBA {} Learned%
}{%
Usman%
\ \protect \BOthers {.}}{%
{\protect \APACyear {2015}}%
}]{%
Usman2015}
\APACinsertmetastar {%
Usman2015}%
\begin{APACrefauthors}%
Usman, S\BPBI M.%
, Jocher, G\BPBI R.%
, Dye, S\BPBI T.%
, McDonough, W\BPBI F.%
\BCBL {}\ \BBA {} Learned, J\BPBI G.%
\end{APACrefauthors}%
\unskip\
\newblock
\APACrefYearMonthDay{2015}{}{}.
\newblock
{\BBOQ}\APACrefatitle {{AGM2015}: Antineutrino {G}lobal {M}ap 2015} {{AGM2015}:
  Antineutrino {G}lobal {M}ap 2015}.{\BBCQ}
\newblock
\APACjournalVolNumPages{Scientific Reports}{5}{1}{13945}.
\newblock
\begin{APACrefDOI} \doi{10.1038/srep13945} \end{APACrefDOI}
\PrintBackRefs{\CurrentBib}

\bibitem [\protect \citeauthoryear {%
Vogel%
\ \BBA {} Beacom%
}{%
Vogel%
\ \BBA {} Beacom%
}{%
{\protect \APACyear {1999}}%
}]{%
vogel1999angular}
\APACinsertmetastar {%
vogel1999angular}%
\begin{APACrefauthors}%
Vogel, P.%
\BCBT {}\ \BBA {} Beacom, J\BPBI F.%
\end{APACrefauthors}%
\unskip\
\newblock
\APACrefYearMonthDay{1999}{}{}.
\newblock
{\BBOQ}\APACrefatitle {Angular distribution of neutron inverse beta decay,
  $\overline\nu_e + p \rightarrow e^+ + n$} {Angular distribution of neutron
  inverse beta decay, $\overline\nu_e + p \rightarrow e^+ + n$}.{\BBCQ}
\newblock
\APACjournalVolNumPages{Physical Review D}{60}{5}{053003}.
\newblock
\begin{APACrefDOI} \doi{10.1103/PhysRevD.60.053003} \end{APACrefDOI}
\PrintBackRefs{\CurrentBib}

\bibitem [\protect \citeauthoryear {%
Wan%
, Hussain%
, Wang%
\BCBL {}\ \BBA {} Chen%
}{%
Wan%
\ \protect \BOthers {.}}{%
{\protect \APACyear {2017}}%
}]{%
Wan:Jinping}
\APACinsertmetastar {%
Wan:Jinping}%
\begin{APACrefauthors}%
Wan, L.%
, Hussain, G.%
, Wang, Z.%
\BCBL {}\ \BBA {} Chen, S.%
\end{APACrefauthors}%
\unskip\
\newblock
\APACrefYearMonthDay{2017}{}{}.
\newblock
{\BBOQ}\APACrefatitle {Geoneutrinos at {Jinping}: {Flux} prediction and
  oscillation analysis} {Geoneutrinos at {Jinping}: {Flux} prediction and
  oscillation analysis}.{\BBCQ}
\newblock
\APACjournalVolNumPages{Phys. Rev. D}{95}{}{053001}.
\newblock
\begin{APACrefDOI} \doi{10.1103/PhysRevD.95.053001} \end{APACrefDOI}
\PrintBackRefs{\CurrentBib}

\bibitem [\protect \citeauthoryear {%
Watanabe%
}{%
Watanabe%
}{%
{\protect \APACyear {2019}}%
}]{%
KAMLAND19}
\APACinsertmetastar {%
KAMLAND19}%
\begin{APACrefauthors}%
Watanabe, H.%
\end{APACrefauthors}%
\unskip\
\newblock
\APACrefYearMonthDay{2019}{}{}.
\newblock
\APACrefbtitle {Geo-neutrino Measurement with {KamLAND}.} {Geo-neutrino
  measurement with {KamLAND}.}
\newblock
\APAChowpublished {Presentation at ``Neutrino Geoscience 2019 Prague'', Czech
  Republic, October 21--23}.
\newblock
\begin{APACrefURL}
  \url{https://indico.cern.ch/event/825708/contributions/3552210/}
  \end{APACrefURL}
\PrintBackRefs{\CurrentBib}

\bibitem [\protect \citeauthoryear {%
Willig%
, Stracke%
, Beier%
\BCBL {}\ \BBA {} Salters%
}{%
Willig%
\ \protect \BOthers {.}}{%
{\protect \APACyear {2020}}%
}]{%
willig2020constraints}
\APACinsertmetastar {%
willig2020constraints}%
\begin{APACrefauthors}%
Willig, M.%
, Stracke, A.%
, Beier, C.%
\BCBL {}\ \BBA {} Salters, V\BPBI J\BPBI M.%
\end{APACrefauthors}%
\unskip\
\newblock
\APACrefYearMonthDay{2020}{}{}.
\newblock
{\BBOQ}\APACrefatitle {{Constraints on mantle evolution from Ce-Nd-Hf isotope
  systematics}} {{Constraints on mantle evolution from Ce-Nd-Hf isotope
  systematics}}.{\BBCQ}
\newblock
\APACjournalVolNumPages{Geochimica et Cosmochimica Acta}{272}{}{36--53}.
\newblock
\begin{APACrefDOI} \doi{10.1016/j.gca.2019.12.029} \end{APACrefDOI}
\PrintBackRefs{\CurrentBib}

\bibitem [\protect \citeauthoryear {%
Wipperfurth%
, Guo%
, {\v S}r{\'a}mek%
\BCBL {}\ \BBA {} McDonough%
}{%
Wipperfurth%
\ \protect \BOthers {.}}{%
{\protect \APACyear {2018}}%
}]{%
wipperfurth:2018}
\APACinsertmetastar {%
wipperfurth:2018}%
\begin{APACrefauthors}%
Wipperfurth, S\BPBI A.%
, Guo, M.%
, {\v S}r{\'a}mek, O.%
\BCBL {}\ \BBA {} McDonough, W\BPBI F.%
\end{APACrefauthors}%
\unskip\
\newblock
\APACrefYearMonthDay{2018}{}{}.
\newblock
{\BBOQ}\APACrefatitle {Earth's chondritic {Th/U}: Negligible fractionation
  during accretion, core formation, and crust--mantle differentiation} {Earth's
  chondritic {Th/U}: Negligible fractionation during accretion, core formation,
  and crust--mantle differentiation}.{\BBCQ}
\newblock
\APACjournalVolNumPages{Earth and Planetary Science Letters}{498}{}{196--202}.
\newblock
\begin{APACrefDOI} \doi{10.1016/j.epsl.2018.06.029} \end{APACrefDOI}
\PrintBackRefs{\CurrentBib}

\bibitem [\protect \citeauthoryear {%
Wipperfurth%
, {\v S}r{\'a}mek%
\BCBL {}\ \BBA {} McDonough%
}{%
Wipperfurth%
\ \protect \BOthers {.}}{%
{\protect \APACyear {2020}}%
}]{%
Wipperfurth2020}
\APACinsertmetastar {%
Wipperfurth2020}%
\begin{APACrefauthors}%
Wipperfurth, S\BPBI A.%
, {\v S}r{\'a}mek, O.%
\BCBL {}\ \BBA {} McDonough, W\BPBI F.%
\end{APACrefauthors}%
\unskip\
\newblock
\APACrefYearMonthDay{2020}{}{}.
\newblock
{\BBOQ}\APACrefatitle {Reference Models for Lithospheric Geoneutrino Signal}
  {Reference models for lithospheric geoneutrino signal}.{\BBCQ}
\newblock
\APACjournalVolNumPages{Journal of Geophysical
  Research}{125}{2}{e2019JB018433}.
\newblock
\begin{APACrefDOI} \doi{10.1029/2019JB018433} \end{APACrefDOI}
\PrintBackRefs{\CurrentBib}

\bibitem [\protect \citeauthoryear {%
Yoshizaki%
\ \BBA {} McDonough%
}{%
Yoshizaki%
\ \BBA {} McDonough%
}{%
{\protect \APACyear {2021}}%
}]{%
yoshizaki2021earth}
\APACinsertmetastar {%
yoshizaki2021earth}%
\begin{APACrefauthors}%
Yoshizaki, T.%
\BCBT {}\ \BBA {} McDonough, W\BPBI F.%
\end{APACrefauthors}%
\unskip\
\newblock
\APACrefYearMonthDay{2021}{}{}.
\newblock
{\BBOQ}\APACrefatitle {{Earth and Mars--distinct inner solar system products}}
  {{Earth and Mars--distinct inner solar system products}}.{\BBCQ}
\newblock
\APACjournalVolNumPages{Geochemistry}{81}{}{125746}.
\newblock
\begin{APACrefDOI} \doi{10.1016/j.chemer.2021.125746} \end{APACrefDOI}
\PrintBackRefs{\CurrentBib}

\end{thebibliography}

\end{document}